\documentclass[aps,prl,notitlepage,twocolumn]{revtex4-1}
\usepackage{graphicx}
\usepackage[colorlinks=true, allcolors=magenta]{hyperref}
\usepackage{xcolor}
\usepackage{amsmath,amssymb,bm}
\usepackage[utf8]{inputenc}
\newcommand{\sectionn}[1]{\underline{\textit{#1}}}
\newcommand{\update}[1]{\textcolor{black}{#1}}

\begin{document}

\title{Antiferromagnet-Based Neuromorphics Using Dynamics of Topological Charges}
\author{Shu Zhang}
\email{suzy@physics.ucla.edu}
\affiliation{Department of Physics and Astronomy, University of California, Los Angeles, California 90095, USA}
\author{Yaroslav Tserkovnyak}
\affiliation{Department of Physics and Astronomy, University of California, Los Angeles, California 90095, USA}
\date{\today}

\begin{abstract}
We propose a spintronics-based hardware implementation of neuromorphic computing, specifically, the spiking neural network, using topological winding textures in one-dimensional antiferromagnets. The consistency of such a network is emphasized in light of the conservation of topological charges, and the natural spatiotemporal interconversions of magnetic winding. We discuss the realization of the leaky integrate-and-fire behavior of neurons and the spike-timing-dependent plasticity of synapses.
\update{Our proposal opens the possibility for an all-spin neuromorphic platform based on antiferromagnetic insulators.}
\end{abstract}

\maketitle

\sectionn{Introduction.}|The concept of neuromorphic computing proposes to combine numerous analog circuits in a network mimicking the architecture of the human brain~\cite{Mead1990}. 
Viewed as a scheme that can outperform the conventional von Neumann machines, especially on complex cognitive tasks, it has undergone intensive development in both learning algorithms and hardware implementations~\cite{Neuromorphics,NeuromorphicsHardware}.

The spiking neuron network (SNN)~\cite{MAASS1997}, among various classes of artificial neuron networks, is so far the closest analog to biological neural systems, due to its unique capability of utilizing the temporal characteristics of events~\cite{tavanaei2019deep,Jang2019SNN}. 
In the SNN, a neuron is a dynamic device featuring the leaky integrate-and-fire (LIF) mechanism~\cite{DELORME1999Spikenet}. Input excitations accumulate on its leaky internal state with a clear threshold, when a spike is triggered. 
Information is then communicated between neurons by these spikes. 
\update{A synapse is a nonvolatile memory with an adaptable weight, which}
controls the amount of influence the firing of the presynaptic neuron has on the postsynaptic one. 
Many SNN learning rules fall under the term spike-timing-dependent plasticity (STDP)~\cite{Hebbian2002,STDPreview}, where the synaptic weight is updated depending on the relative spiking time of the pre- and postsynaptic neurons within a 
time window.

Because SNNs can be both computationally more efficient and less energy consuming, their hardware implementations have attracted a lot of interest. 
Spintronics, the study of intrinsically nonlinear spin dynamics and transport,
has 
demonstrated to be promising for implementing neuromorphic computing~\cite{Grollier2016review,sengupta2018neuromorphic,UMESH2019survey,grollier2020neuromorphic,KurenkovReview}. 
Many nanoscale spintronic devices, with advantages in terms of nonvolatile memory, fast switching, high endurance and low energy dissipation, have already been developed for memory and logic applications~\cite{Allwood2005,parkin2008DW,Fert2008review,fert2013skyrmions,Bhatti2017RAM}. Various candidate neuron- and/or synapse-like \update{devices to replace elements in the traditional CMOS circuitry} have been proposed based on magnetic tunnel junctions~\cite{Krzysteczko2012MTJ,Sengupta2015MTJ,Vincent2015MTJ,Zhang2016MTJ,Zhang2016MTJ2,torrejon2017MTJ,kurenkov2019}, spin torque driven nano-oscillators~\cite{Khymyn2018ST,Sulymenko2019Logic,Fukami2018ST}, domain-wall~\cite{Yue2019DW,Sharad2013DW,Cui2019} and skyrmion-bubble~\cite{Huang2017Skyr,Li2017Skyr,Chen2018Skyr} racetracks, and spin-wave dynamics~\cite{Zeng2016SW,Katayama2016SW}.

In this Letter, we propose an SNN scheme based on dynamics of topological winding textures. Our proposal emphasizes physical compatibility of neurons and synapses, which is crucial to form a scalable and efficient network~\cite{kurenkov2019}.
\update{Particularly, the proposed SNN would, in its ideal form,
operate relying purely on spin dynamics. Spiking signals flow continuously throughout the network, their generation and transmission following the same spin-dynamical principles.} 
By enhancing the physical integrity of the network, we can minimize energy cost and simplify operational rules.

\begin{figure*}[ht]
    \centering
    \includegraphics[width = \linewidth]{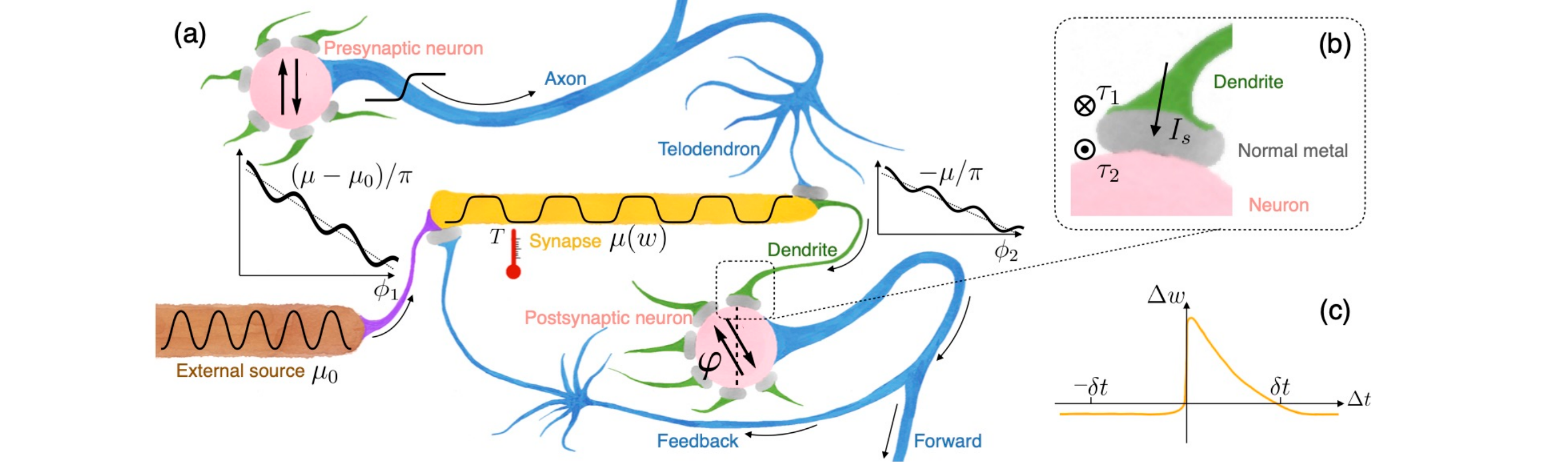}
    \caption{
    Schematics of neurons and a synapse in an SNN based on dynamics of topological charges in biaxial antiferromagnets. 
    (a) A neuron is a single-domain magnet with angular order parameter $\varphi$. The neuron fires when $\varphi$ rotates (counterclockwise) by $\pi$, creating a topological charge with $\zeta = + 1$. 
    An axon branches out into a feedback arm and a forward one, each further fans out into several telodendrons.
    Arrows indicate directions of signal flow. 
    A synapse is a nanostrip holding a winding texture, a thermodynamic system (with temperature $T$ and chemical potential $\mu$) of interacting topological charges, their net density $w$ as the synaptic weight. 
    An external source of winding with constant chemical potential $\mu_0$ is used to increase the synaptic weight $w$ upon causal firings.
    Sketches of $\cos\phi(x)$ inside the synapse and the external source indicate the low and high density of topological charges.
    The order parameters $\phi_1$ and $\phi_2$ at the two ends of the synapse experience tilted washboard potentials.
    (b) A weak coupling between the neuron and the dendrite mediated by a metal spacer.
    A topological charge arrives from the dendrite and pumps a spin current $I_s$ into the normal metal, exerting torques of opposite signs on the two interfaces.
    (c) The learning function \update{(the change in the synaptic weight $\Delta w$ versus the relative arrival time $\Delta t = t_\text{post} - t_\text{pre}$ of the post- and presynaptic signals)} of a possible scheme of STDP of the synapse. Causal firings strengthen the synaptic connection.
    }
    \label{fig:scheme}
\end{figure*}

To illustrate our idea, 
we consider quasi-one-dimensional antiferromagnets with an easy-plane anisotropy.
With the N{\'e}el order 
$\mathbf{n} =(\cos \phi, \sin \phi)$ 
constrained in the easy plane, its azimuthal angle $\phi (x,t)$ parametrizes the local magnetic state. 
Even before accounting for any energetics, we 
recognize a \textit{topological} conservation law~\cite{TopoConservation,NutshellQFT}:
\begin{equation}
    \partial_t \rho + \partial_x j = 0,
    \label{eq:topological-conservation}
\end{equation} 
where 
$\rho = - \partial_x \phi/\pi$ and 
$j = \partial_t \phi/\pi$
are the topological charge density and current. 
In real materials, the rotational symmetry within the easy plane is usually reduced by anisotropies. We thus consider the Hamiltonian density~\cite{Baltz2018AFreview}:
\begin{equation}
    \mathcal{H} = 
    \frac{\mathcal{A}}{2} (\partial_x \phi)^2 
    + \frac{m^2}{2 \chi}
    + \frac{\mathcal{K}}{2} \sin^2 \phi,
    \label{eq:hamiltonian}
\end{equation}
where $\mathcal{A}$ is the exchange stiffness, $\chi$ and $m$ are the spin susceptibility and the spin density (per unit length) transverse to the easy plane, and $\mathcal{K} > 0$ is an axial in-plane anisotropy.

Equation~(\ref{eq:topological-conservation}) reflects the robust nature of a topological winding texture in one dimension, which is made up of chiral domain walls.
A domain wall interpolates between the two ground states ($\phi = 0$ or $\pi$) of the Hamiltonian~(\ref{eq:hamiltonian}), and possesses a topological charge 
$\zeta = \int dx \, \rho = \pm 1$, with the sign designating chirality.
The net total topological charge of a magnetic strip is a conserved quantity. The continuity equation~(\ref{eq:topological-conservation}) also reveals the dynamical aspects of the winding texture. The motion of a domain wall through a spatial point $x_0$ necessarily involves a $\pi$ winding of $\phi(x_0, t)$ in time, and vice versa. There thus exists a natural spatiotemporal interconversion of topological winding.

In our proposal, hydrodynamics of topological charges governs the information transmission and processing in the entire network. As shown in Fig.~\ref{fig:scheme}(a), all elements, excepts for a few links, are made of antiferromagnetic materials.
Axons and dendrites are 
\update{channels for ballistic}
propagation of topological charges. 
The neuronal behavior is realized by a single-domain magnet with leaky (damped) dynamics. Its internal state, described by the uniform order parameter $\varphi(t)$, is analog, while its output is digital: A domain wall is created and propagates into the axon, if $\varphi$ winds by $\pi$, or not, if $\varphi$ relaxes to its initial ground state. 
A synapse contains a magnetic nanostrip holding a winding texture.
The net density of topological charges serves as a nonvolatile analog memory of the synaptic weight $w$, which can be adjusted according to the relative arrival time $\Delta t$ of the 
domain walls fired by the pre- and postsynaptic neurons. 
In this way, neurons and synapses can be naturally integrated into a 
compatible network,
driven solely by the continuous flow of topological charges. 

\sectionn{Hydrodynamics of topological charges.}|We now study the dynamics governed by the Hamiltonian~(\ref{eq:hamiltonian}) 
together with the Rayleigh dissipation function density
$\mathcal{R} \!=\! ( \alpha \mathcal{J} / 2) (\partial_t \phi)^2$, where $\alpha$ is the dimensionless Gilbert damping~\cite{Gilbert2004} and $\mathcal{J}$ is the saturated spin density. 
The in-plane angle $\phi$ and the transverse spin (angular momentum) density $m$ are conjugate variables, satisfying the Poisson bracket 
$\{ \phi(x),  m(x') \} \!=\!\delta (x-x')$.
Equations of motion are thus given by the generalized Hamilton equations
$\partial_t \phi \!=\!\partial \mathcal{H} / \partial m$, and 
$\partial_t m\!=\! -\partial \mathcal{H} / \partial \phi  - \partial \mathcal{R} / \partial (\partial_t \phi)$.
Eliminating $m$ yields the damped sine-Gordon equation~\cite{Tserkovnyak2018Perspective,Hill2018LJJ}
$
    (1/u^2)\partial_t^2 \phi - \partial_x^2 \phi + (1/2 \lambda^2) \sin 2 \phi + \beta \partial_t \phi \!=\! 0,
$
where \update {the spin-wave speed 
$ u \!=\! \sqrt{\mathcal{A} /\chi}$, the domain-wall width $\lambda =\sqrt{\mathcal{A}/\mathcal{K}}$} and the damping factor 
$\beta \!=\! \alpha \mathcal{J}  / \mathcal{A}$. 
We consider its solution in the low-damping limit $\beta \rightarrow 0$ (to recover Lorentz invariance)~\cite{khymyn2017,Haldane1983Lorentz}, 
\begin{equation}
    \phi_0 (x,t) = 2 \arctan \left[ \exp \left( - \zeta \frac{x-vt}{\lambda \sqrt{1-v^2/u^2}} \right)\right],
    \label{eq:single-soliton}
\end{equation}
which \update{is often referred to as a soliton. It} describes a domain wall with topological charge $\zeta  \!=\!  \pm 1$ moving with speed $v$. When topological charges are far apart, their interaction is exponentially weak and they behave like free particles. \update{A small damping} can be accounted for by a weak frictional force~\update{\cite{Nonlinear1980}}, resulting the energy dissipation 
$d \mathcal{H} / dt  \!=\!  -2 \mathcal{R}$, which translates into 
$dv/dt = -\beta v(u^2-v^2)$.
By tuning the size of the in-plane anisotropy $\mathcal{K}$, domain walls can be sharp, just like spikes in biological nerves.
The solitonic motion of topological charges in our proposal bears resemblance to the nonlinear action-potential propagation in the Hodgkin-Huxley model~\cite{hodgkin1952propagation}.

As we have pointed out, the hydrodynamics of topological charges is rooted in the conservation law~(\ref{eq:topological-conservation}). 
\update{A large energy barrier, determined by the easy-plane anisotropy $\mathcal{K}_z$ ($\gg \mathcal{K}$), must be overcome for thermally activated phase slips and (for very small structures and cryogenic temperatures) quantum tunnelings~\cite{Kim2016PhaseSlip,Ivanov1998} to relax the metastable winding textures. To stay in the valid regime of the conservation law, we operate at} 
temperatures much below the N{\'e}el temperature, and \update{with} the density of domain walls much below the Landau criterion for their energetic stability
($\rho \ll 1/\sqrt{\mathcal{A}/\mathcal{K}_z}$~\cite{Sonin2010Review}).

The topological conservation law guarantees the transmission of a topological charge through a strong coupling: an interface between two magnets in direct contact through magnetic exchange,
\update{which is used between a neuron and its axon, and between an axon and its telodendrons.} The signal traveling along an axon fans out as a topological charge enters each telodendron, albeit in practice there may need to be some amplification to compensate for cascading and losses of the energy.

\sectionn{Neuron.}|For a neuron to be able to integrate topological charges from many of its dendrites, however, we need a weak coupling, as sketched in Fig.~\ref{fig:scheme}(b), in order to transfer spatial windings in dendrites into (weak) temporal stimuli for the neuron, as well as impedes the feedback to dendrites when the neuron fires.
The  weak  coupling  can be  realized  by  inserting  a  nonmagnetic-metal  spacer  between  the  neuron  and  dendrites,  forming a spin-valve-like structure~\cite{Tserkovnyak2003SpinPump}.

Since a topological charge is ``quantized,'' its transmission from the dendrite to the metal spacer (as a spin-injection pulse, as detailed below) is a question of yes or no.
Invoking the boundary condition~\cite{Hill2018LJJ}
$A \partial_x \phi - \tau_1 = 0$,
a domain wall~(\ref{eq:single-soliton}) with speed 
$v = 8 \pi \mathcal{A} / \hbar g$ can slip through without any deformation, which is reminiscent of the microwave impedance matching for optimal transmission. There must therefore exist a finite speed window for the topological charges so that signals reach the neuron.

The neuron in our proposal is a single-domain antiferromagnet with a uniform magnetic state 
$\partial_x \varphi  \!=\!  0$, where $\varphi$ is the azimuthal angle of its N{\'e}el order parameter.
We restate the Hamiltonian 
$\widetilde{H} = M^2/2\widetilde{\chi}+(\widetilde{\mathcal{K}}/2)\sin^2 \varphi$ and the Rayleigh dissipation function $\widetilde{R} = (\widetilde{\alpha} \widetilde{\mathcal{J}}/2) (\partial_t \varphi)^2$, where the tilded parameters are defined as previously but for a zero-dimensional system, the neuron.
The azimuthal angle $\varphi$ and the transverse angular momentum $M$ 
are canonical conjugate variables. We obtain equations of motion~\cite{khymyn2017}
\begin{equation}
    \partial_t \varphi = \frac{M}{\widetilde{\chi}}, \text{ and }
    \partial_t M = - \frac{\widetilde{\mathcal{K}}}{2} \sin 2\varphi - \widetilde{\alpha} \widetilde{\mathcal{J}} \partial_t \varphi + \tau(t),
    \label{eq:neuron}
\end{equation}
where the torque $\tau(t)$ is 
a result of incoming signals from the dendrite, which we now turn to. 

When a solitonic domain wall \update{(with positive chirality)} propagating along the dendrite unwinds at the dendrite-metal interface, the spin dynamics pumps a spin current 
$I_s \!=\! (\hbar g / 4\pi) \partial_t \phi$ 
into the normal metal~\cite{Tserkovnyak2003SpinPump} 
[note $I_s \!\propto\! j$, see Eq.~(\ref{eq:topological-conservation})], where $g$ is the spin-mixing conductance. 
Consider the simplified situation where the spin current partitions into half transmission and half reflection~\cite{supmat}.
A spin current is an angular momentum flux. In other words, a torque is exerted at the metal-dendrite interface, and equally at the metal-neuron interface~\cite{Heinrich2003SpinPump},
$-\tau_1 \!=\! \tau_2 \!=\!  I_s/2 = (\hbar g/ 8\pi) \partial_t \phi$. 
The total angular-momentum transfer into the neuron is thus
$\delta M \!=\! \int dt \tau_2\!=\! \hbar g/ 8$.

\begin{figure}[t]
    \includegraphics[width = \linewidth]{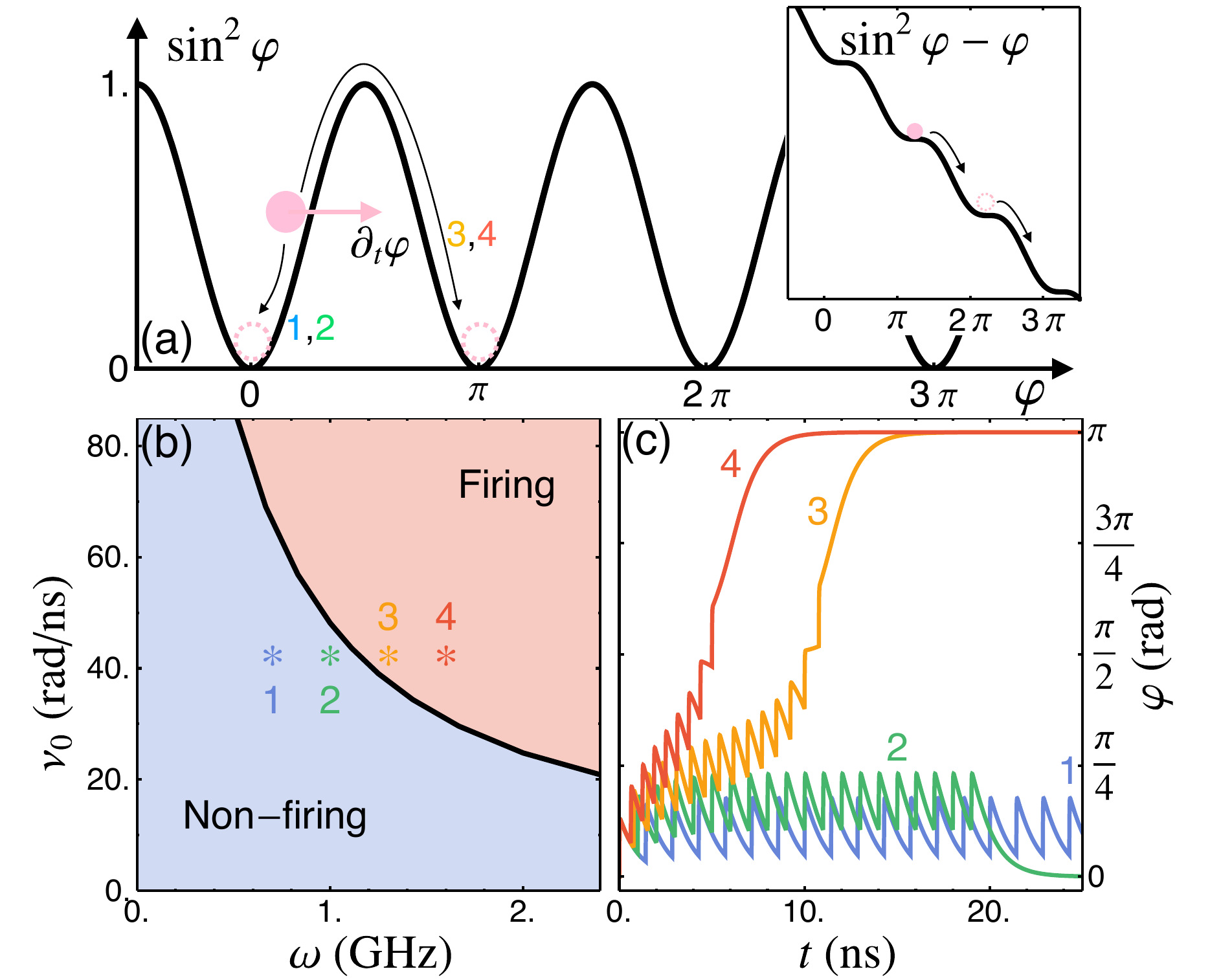} 
    \caption{Leaky integrate-and-fire functionality of the neuron. 
    (a) The washboard potential. The angular order parameter $\varphi$ of the neuron and its time derivative $\partial_t \varphi$ are analogous to the position and speed of a particle. Inset: The tilted washboard potential where local minima become unstable. 
    (b) The firing and nonfiring diagram depending on the input rate $\omega$ and amplitude $\nu_0$ of the incoming train of angular momentum kicks. Numerics are done with parameters $\omega_0 = 10$~GHz and $\eta = 5$.
    (c) The time evolution of the order parameter of the neuron $\varphi(t)$ simulated for samples marked in the diagram. In each simulation, 20 pulses are sent to the neuron until it fires.}
    \label{fig:neuron}
\end{figure}

Eliminating $M$ in Eq.~(\ref{eq:neuron}),
$\widetilde{\chi}\partial_t^2 \varphi \!+\! (\widetilde{\mathcal{K}}/2)\sin 2 \varphi \!+\! \widetilde{\alpha} \widetilde{\mathcal{J}} \partial_t \varphi  \!=\!  \tau(t)$ describes a familiar driven-and-damped particle in a washboard potential with local minima $\varphi  \!=\!  n \pi$ ($n \in \mathbb{Z}$), as shown in Fig.~\ref{fig:neuron}(a). A few characteristic parameters can be extracted by taking the approximation to a damped harmonic oscillator near the local minimum: the undamped frequency 
$\omega_0  \!=\!  \sqrt{\widetilde{\mathcal{K}}/\widetilde{\chi}}$ and the damping ratio 
$\eta = \widetilde{\alpha}\widetilde{\mathcal{J}} \Big{/} 2\sqrt{\widetilde{\mathcal{K}} \widetilde{\chi}}$.
We set the material parameters in the overdamped regime with
$\eta \!>\! 1$, the relaxation time 
$t_0  \!=\!  1/(\eta - \sqrt{\eta^2-1}) \omega_0$
defines the local timescale on which the neuron performs integration. 
Compared with the slow reaction of the neuron, the angular-momentum transfer due to a sharp incoming domain wall can be almost instantaneous\update{~\cite{supmat}}.
Thus $\tau(t)$ can be regarded as composed of delta pulses, each elevates the canonical momentum $M$ by $\delta M$, and, as manifested by Eq.~(\ref{eq:neuron}), boosts $\partial_t\varphi$ by $\nu_0 \!=\! \delta M / \widetilde{\chi}$.

Take the example of an initially stationary state trapped at 
$\varphi|_{t=0} \!=\! 0$, a sudden angular momentum kick sets the initial condition 
$\partial_t \varphi|_{t=0^+} \!=\! \nu_0$ for the relaxation process. 
As a train of topological charges coming in at a rate $\omega$, the state $\varphi$ wiggles gently as the effect of each kick fades, featuring a leaky integration. See Fig.~\ref{fig:neuron}(c). Only with a sufficiently high input rate and/or large amplitude, can the signal pulses trigger the neuron to fire---it 
overcomes the energy barrier and falls into the neighboring local minimum 
$\varphi  \!=\!  \pi$. The dependence of the firing or nonfiring result on the rate and amplitude of pulses is plotted in Fig.~\ref{fig:neuron}(b). 
When the neuron fires, it experiences a phase jump of $\pi$, whereas the phase deep in the axon remains unchanged. A domain wall with topological charge $\zeta  \!=\!  +1$ is thus created on a timescale of $t_0$, entering the axon with a certain initial velocity.

Another behavior of a biological neuron that we can imitate is the bursting\update{---}a neuron fires repeatedly in groups separated by quiescent periods~\cite{IZHIKEVICH2000bursting}. For our artificial neuron, a constant torque $\tau_0$ (such as a spin transfer torque used in Ref.~\cite{Sulymenko2019Logic}) effectively tilts the washboard potential. This can cause repetitive firing once the torque exceeds a threshold where the local minima become unstable. On average, the firing rate is proportional to the tilting,
$r = \tau_0/\pi\widetilde{\alpha}\widetilde{\mathcal{J}}$.

\sectionn{Synapse.}|A nonvolatile analog memory with an adjustable weight, as the key element of a synapse, is realized in our proposal by an antiferromagnetic nanostrip~(\ref{eq:hamiltonian}) holding a metastable winding texture. We define the dimensionless density of topological charges $w = N \lambda/L$ as the synaptic weight, where $N$ is the net number of topological charges and $L$ is the length of the nanostrip, which is large enough ($L \gg \lambda$) to allow us to consider $w$ as an essentially continuous variable.
The nanostrip is a one-dimensional thermodynamic system of interacting topological charges.
\update{The interaction gives rise to a density dependent chemical potential}
$\mu (w)= \delta F / \delta N$, where $F$ is its total free energy. 

\update{Here, we outline a proof-of-principle proposal for the STDP procedure, which can be modified and improved, subject to practical experimental considerations. We update the synaptic weight in two ways: an increase upon causal firings, and a constant degradation background when the causal relation is not maintained. See Fig.~\ref{fig:scheme}(c).} 

\begin{figure}[b]
    \centering
    \includegraphics[width = \linewidth]{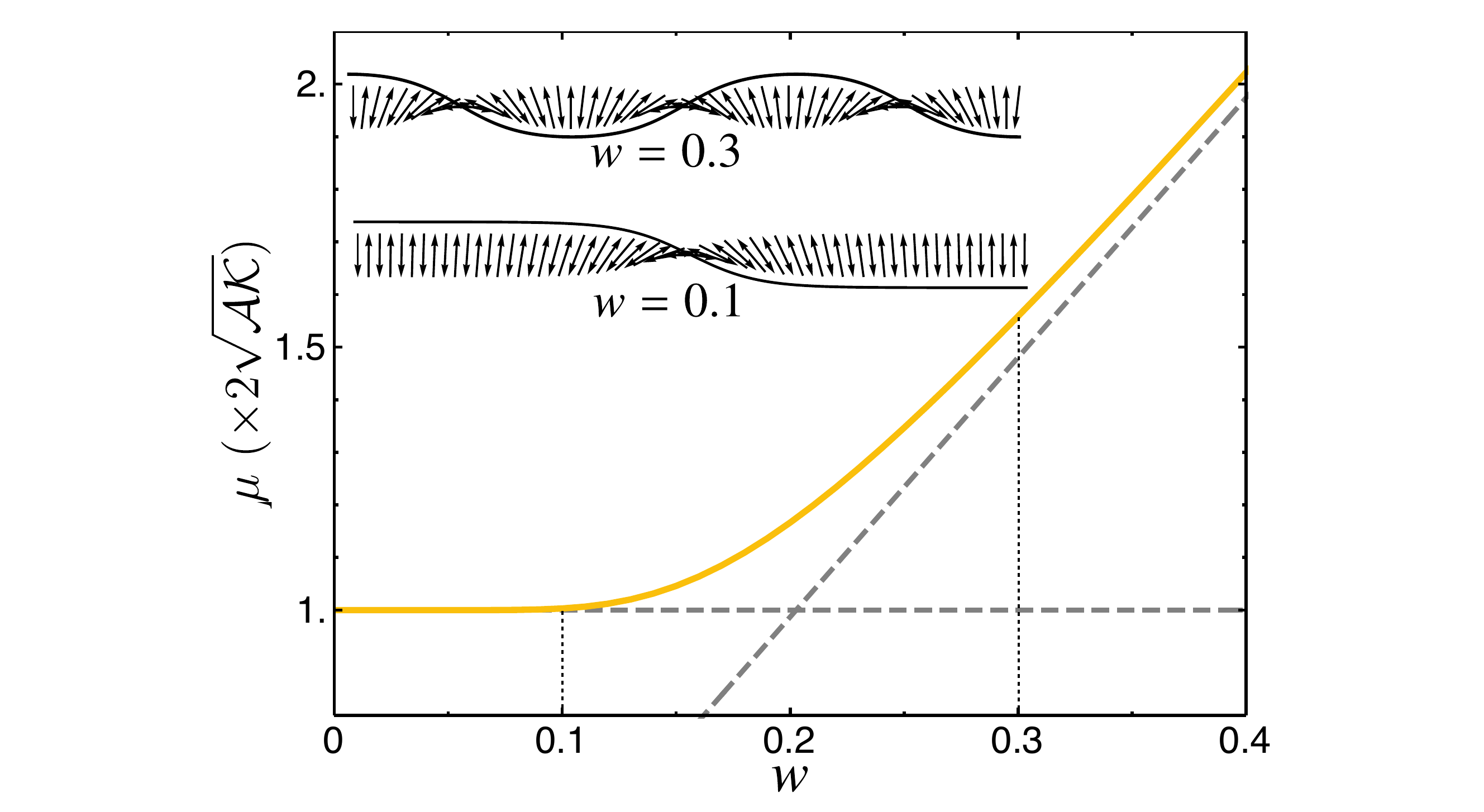}
    \caption{The nonlinear dependence of the chemical potential $\mu$ on the synaptic weight $w$, calculated for the metastable ground-state textures of the Hamiltonian~(\ref{eq:hamiltonian}), which are the solutions to the static sine-Gordon equation~\cite{supmat}. Inset: the magnetic textures and their $\cos \phi(x)$ for two values of $w$. The texture becomes closer to a uniform spiral as $w$ approaches 1. The asymptote has the analytical expression 
    $\mu = \sqrt{\mathcal{A}\mathcal{K}} \pi^2 w$. }
    \label{fig:synapse}
\end{figure}

\update{We first consider the latter, focusing on the right end of the synapse in Fig.~\ref{fig:scheme}(a).
The local parameter $\phi_2$, if unperturbed, stays at a local minimum of a tilted washboard potential due to the in-plane anisotropy. The tilting $-\mu/\pi$ is present since an amount of energy $\mu$ is needed to  inject an extra topological charge into the synapse through winding $\phi_2$ by $\pi$.}
\update{Whenever a topological charge from the presynaptic neuron arrives, the dynamics of $\phi_2$ is activated: $\phi_2$ is brought out of the local minimum either as a result of kinetic perturbation, or heating, which softens the anisotropy. Both can be due to the spin dynamics associated with the incoming topological charge.}
\update{The synapse is therefore open briefly before damping sends $\phi_2$ back to a local equilibrium.}
A small amount of topological charges are pushed into the dendrite, passing a modulated signal to the postsynaptic neuron.
\update{The strength of this signal is determined by the synaptic weight: The denser the topological charges in the synapse, the higher the chemical potential (see Fig.~\ref{fig:synapse}), so that more topological charges are pushed into the dendrite, like a compressed gas being briefly released.} 

\update{An external reservoir of topological charges (a ``chirality battery''~\cite{Tserkovnyak2018Battery}), which is maintained at a constant chemical potential $\mu_0 > \mu$, is linked to the left end of the synapse, where the local order parameter $\phi_1$, similar to the situation with $\phi_2$, experiences a tilted washboard potential with an average slope proportional to the relative chemical potential $\mu - \mu_0$. See Fig.~\ref{fig:scheme}(a).
We now wish 
the combination of the presynaptic signal and the causal postsynaptic feedback (if created) to increase the synaptic weight. We consider one possibility following Ref.~\cite{kurenkov2019}. The kinetic perturbation caused by the postsynaptic signal can activate the dynamics of $\phi_1$ only when assisted by the remnant heat from the presynaptic signal, so that topological charges flow from the external reservoir into the synapse, where the magnetic texture is naturally damped into a new equilibrium with higher $\mu$.}
The STDP time window $\delta t$ is determined by the \update{heat transport properties} of the nanostrip.
\update{Other, nonheat-based scenarios can also be envisioned, which we leave for a future work.}

\sectionn{Discussion.}|As a route to cognition mimicking, SNNs are by nature complicated and approximative. 
Magnetic materials in different parameter regimes 
have to be \update{tailored} to realize functionalities in different elements, \update{which can be made possible by the large range of insulating antiferromagnetic materials at disposal~\cite{Tserkovnyak2005Review}.} Antiferromagnets are free of stray fields, thus allowing for the dense packing of devices. They also have fast spin dynamics, with the bulk magnon gap often on the THz scale. 
\update{Candidate materials can be antiferromagnets relevant to spin superfluidity~\cite{superfluidCrO}, such as cubic perovskites~\cite{Takei2014AFM}, as the winding dynamics would benefit from low crystalline anisotropies, or a versatile platform of magnetic van der Waals materials~\cite{vanderWaals}, or synthetic antiferromagnets, which already found a promising application in racetrack memory~\cite{syntheticAFM}. }
\update{The operating frequency could reach the GHz range, estimated from the characteristic timescale of the neuron $t_0 \sim 1$~ns using $\widetilde{\mathcal{K}} \sim 10^{-22}$~J, $\widetilde{\chi} \sim 10^{-42}$~J$\cdot$s$^2$, $\widetilde{J} \sim 10^5 \hbar$ for a neuron of volume $\sim 10^3$~(nm)$^3$, and $\widetilde{\alpha} \sim 10^{-2}$.}
\update{Due to the all-spin design, the energy efficiency of our proposal can be very compelling. Once we put the free energy associated with magnetic winding into the system, collective spin dynamics merely redistribute the energy by shuffling the winding in different parts, apart from the energy losses due to Gilbert damping and effective phase slips in the junctions. Two main energy-costing processes involve the firing of the neuron and the  propagation of domain walls in the conduits. The neuron has a very small energy barrier $\sim 10$~zJ, though practically monodomain switching can cost, for example, $\sim 0.1$~pJ in a magnetic tunnel junction. 
To operate at GHz frequency, solitonic domain walls would ideally be sent at a speed $v \sim 1$~km/s along axons of $\mu$m scale. Integrating the Rayleigh function over the size of the domain wall $\lambda$, we obtain an energy dissipation rate $\sim \alpha \mathcal{J} v^2 /2 \lambda \sim 1$ pJ/s by taking $\alpha \sim 10^{-4}$, $\mathcal{J} \sim 10^3 \hbar$/nm and $\lambda \sim 10$~nm. The energy cost of each spike transmission is therefore negligible ($\sim 1$~zJ). 
}

\update{In summary, our proposal has exploited the conservation law of topological charges and the conversion between temporal and spatial properties of magnetic topological textures to provide a unified ground for dynamical integration, signal propagation, and nonvolatile memory.}
This principle may also be applied to SNN schemes involving other types of topological charges, such as vortices in two dimensions and hedgehogs in three dimensions. 
\phantom{\cite{Tserkovnyak2017Boundary,Elliptic}}

\begin{acknowledgments}
    We especially thank Aleksandr Kurenkov for sharing his knowledge on neuromorphics. We thank Robijn Bruinsma, Se Kwon Kim, Mayank Mehta, \update{Jiang Xiao}, Giovanni Zocchi and Ji Zou for useful discussions. \update{We would also like to thank the anonymous referees for their valuable suggestions to improve the manuscript.} This work is supported by NSF under Grant No.~DMR-1742928.
\end{acknowledgments}

\bibliography{main}

\begin{thebibliography}{62}%
\makeatletter
\providecommand \@ifxundefined [1]{%
 \@ifx{#1\undefined}
}%
\providecommand \@ifnum [1]{%
 \ifnum #1\expandafter \@firstoftwo
 \else \expandafter \@secondoftwo
 \fi
}%
\providecommand \@ifx [1]{%
 \ifx #1\expandafter \@firstoftwo
 \else \expandafter \@secondoftwo
 \fi
}%
\providecommand \natexlab [1]{#1}%
\providecommand \enquote  [1]{``#1''}%
\providecommand \bibnamefont  [1]{#1}%
\providecommand \bibfnamefont [1]{#1}%
\providecommand \citenamefont [1]{#1}%
\providecommand \href@noop [0]{\@secondoftwo}%
\providecommand \href [0]{\begingroup \@sanitize@url \@href}%
\providecommand \@href[1]{\@@startlink{#1}\@@href}%
\providecommand \@@href[1]{\endgroup#1\@@endlink}%
\providecommand \@sanitize@url [0]{\catcode `\\12\catcode `\$12\catcode
  `\&12\catcode `\#12\catcode `\^12\catcode `\_12\catcode `\%12\relax}%
\providecommand \@@startlink[1]{}%
\providecommand \@@endlink[0]{}%
\providecommand \url  [0]{\begingroup\@sanitize@url \@url }%
\providecommand \@url [1]{\endgroup\@href {#1}{\urlprefix }}%
\providecommand \urlprefix  [0]{URL }%
\providecommand \Eprint [0]{\href }%
\providecommand \doibase [0]{http://dx.doi.org/}%
\providecommand \selectlanguage [0]{\@gobble}%
\providecommand \bibinfo  [0]{\@secondoftwo}%
\providecommand \bibfield  [0]{\@secondoftwo}%
\providecommand \translation [1]{[#1]}%
\providecommand \BibitemOpen [0]{}%
\providecommand \bibitemStop [0]{}%
\providecommand \bibitemNoStop [0]{.\EOS\space}%
\providecommand \EOS [0]{\spacefactor3000\relax}%
\providecommand \BibitemShut  [1]{\csname bibitem#1\endcsname}%
\let\auto@bib@innerbib\@empty
\bibitem [{\citenamefont {{Mead}}(1990)}]{Mead1990}%
  \BibitemOpen
  \bibfield  {author} {\bibinfo {author} {\bibfnamefont {C.}~\bibnamefont
  {{Mead}}},\ }\href {\doibase 10.1109/5.58356} {\bibfield  {journal} {\bibinfo
   {journal} {Proc. IEEE}\ }\textbf {\bibinfo {volume} {78}},\ \bibinfo {pages}
  {1629} (\bibinfo {year} {1990})}\BibitemShut {NoStop}%
\bibitem [{\citenamefont {{Arbib}}\ and\ \citenamefont
  {{Bonaiuto}}(2016)}]{Neuromorphics}%
  \BibitemOpen
  \bibfield  {author} {\bibinfo {author} {\bibfnamefont {M.~A.}\ \bibnamefont
  {{Arbib}}}\ and\ \bibinfo {author} {\bibfnamefont {J.~J.}\ \bibnamefont
  {{Bonaiuto}}},\ }\href@noop {} {\emph {\bibinfo {title} {From Neuron to
  Cognition via Computational Neuroscience}}}\ (\bibinfo  {publisher} {MIT
  Press},\ \bibinfo {address} {Cambridge, MA},\ \bibinfo {year}
  {2016})\BibitemShut {NoStop}%
\bibitem [{\citenamefont {{Suri}}(2017)}]{NeuromorphicsHardware}%
  \BibitemOpen
  \bibfield  {author} {\bibinfo {author} {\bibfnamefont {M.}~\bibnamefont
  {{Suri}}},\ }\href@noop {} {\emph {\bibinfo {title} {Advances in Neuromorphic
  Hardware Exploiting Emerging Nanoscale Devices}}}\ (\bibinfo  {publisher}
  {Springer Nature},\ \bibinfo {address} {India},\ \bibinfo {year}
  {2017})\BibitemShut {NoStop}%
\bibitem [{\citenamefont {Maass}(1997)}]{MAASS1997}%
  \BibitemOpen
  \bibfield  {author} {\bibinfo {author} {\bibfnamefont {W.}~\bibnamefont
  {Maass}},\ }\href {\doibase https://doi.org/10.1016/S0893-6080(97)00011-7}
  {\bibfield  {journal} {\bibinfo  {journal} {Neural Networks}\ }\textbf
  {\bibinfo {volume} {10}},\ \bibinfo {pages} {1659 } (\bibinfo {year}
  {1997})}\BibitemShut {NoStop}%
\bibitem [{\citenamefont {Tavanaei}\ \emph {et~al.}(2019)\citenamefont
  {Tavanaei}, \citenamefont {Ghodrati}, \citenamefont {Kheradpisheh},
  \citenamefont {Masquelier},\ and\ \citenamefont {Maida}}]{tavanaei2019deep}%
  \BibitemOpen
  \bibfield  {author} {\bibinfo {author} {\bibfnamefont {A.}~\bibnamefont
  {Tavanaei}}, \bibinfo {author} {\bibfnamefont {M.}~\bibnamefont {Ghodrati}},
  \bibinfo {author} {\bibfnamefont {S.~R.}\ \bibnamefont {Kheradpisheh}},
  \bibinfo {author} {\bibfnamefont {T.}~\bibnamefont {Masquelier}}, \ and\
  \bibinfo {author} {\bibfnamefont {A.}~\bibnamefont {Maida}},\ }\href
  {\doibase https://doi.org/10.1016/j.neunet.2018.12.002} {\bibfield  {journal}
  {\bibinfo  {journal} {Neural Networks}\ }\textbf {\bibinfo {volume} {111}},\
  \bibinfo {pages} {47} (\bibinfo {year} {2019})}\BibitemShut {NoStop}%
\bibitem [{\citenamefont {{Jang}}\ \emph {et~al.}(2019)\citenamefont {{Jang}},
  \citenamefont {{Simeone}}, \citenamefont {{Gardner}},\ and\ \citenamefont
  {{Gruning}}}]{Jang2019SNN}%
  \BibitemOpen
  \bibfield  {author} {\bibinfo {author} {\bibfnamefont {H.}~\bibnamefont
  {{Jang}}}, \bibinfo {author} {\bibfnamefont {O.}~\bibnamefont {{Simeone}}},
  \bibinfo {author} {\bibfnamefont {B.}~\bibnamefont {{Gardner}}}, \ and\
  \bibinfo {author} {\bibfnamefont {A.}~\bibnamefont {{Gruning}}},\ }\href
  {\doibase 10.1109/MSP.2019.2935234} {\bibfield  {journal} {\bibinfo
  {journal} {IEEE Sig. Process. Mag.}\ }\textbf {\bibinfo {volume} {36}},\
  \bibinfo {pages} {64} (\bibinfo {year} {2019})}\BibitemShut {NoStop}%
\bibitem [{\citenamefont {{Delorme}}\ \emph {et~al.}(1999)\citenamefont
  {{Delorme}}, \citenamefont {{Gautrais}}, \citenamefont {{van Rullen}},\ and\
  \citenamefont {{Thorpe}}}]{DELORME1999Spikenet}%
  \BibitemOpen
  \bibfield  {author} {\bibinfo {author} {\bibfnamefont {A.}~\bibnamefont
  {{Delorme}}}, \bibinfo {author} {\bibfnamefont {J.}~\bibnamefont
  {{Gautrais}}}, \bibinfo {author} {\bibfnamefont {R.}~\bibnamefont {{van
  Rullen}}}, \ and\ \bibinfo {author} {\bibfnamefont {S.}~\bibnamefont
  {{Thorpe}}},\ }\href {\doibase https://doi.org/10.1016/S0925-2312(99)00095-8}
  {\bibfield  {journal} {\bibinfo  {journal} {Neurocomputing}\ }\textbf
  {\bibinfo {volume} {26-27}},\ \bibinfo {pages} {989 } (\bibinfo {year}
  {1999})}\BibitemShut {NoStop}%
\bibitem [{\citenamefont {Gerstner}\ and\ \citenamefont
  {Kistler}(2002)}]{Hebbian2002}%
  \BibitemOpen
  \bibfield  {author} {\bibinfo {author} {\bibfnamefont {W.}~\bibnamefont
  {Gerstner}}\ and\ \bibinfo {author} {\bibfnamefont {W.~M.}\ \bibnamefont
  {Kistler}},\ }\href
  {https://link.springer.com/article/10.1007/s00422-002-0353-y} {\bibfield
  {journal} {\bibinfo  {journal} {Biol. Cybernet.}\ }\textbf {\bibinfo {volume}
  {87}},\ \bibinfo {pages} {404} (\bibinfo {year} {2002})}\BibitemShut
  {NoStop}%
\bibitem [{\citenamefont {Caporale}\ and\ \citenamefont
  {Dan}(2008)}]{STDPreview}%
  \BibitemOpen
  \bibfield  {author} {\bibinfo {author} {\bibfnamefont {N.}~\bibnamefont
  {Caporale}}\ and\ \bibinfo {author} {\bibfnamefont {Y.}~\bibnamefont {Dan}},\
  }\href {\doibase 10.1146/annurev.neuro.31.060407.125639} {\bibfield
  {journal} {\bibinfo  {journal} {Annu. Rev. Neurosci.}\ }\textbf {\bibinfo
  {volume} {31}},\ \bibinfo {pages} {25} (\bibinfo {year} {2008})}\BibitemShut
  {NoStop}%
\bibitem [{\citenamefont {{Grollier}}\ \emph {et~al.}(2016)\citenamefont
  {{Grollier}}, \citenamefont {{Querlioz}},\ and\ \citenamefont
  {{Stiles}}}]{Grollier2016review}%
  \BibitemOpen
  \bibfield  {author} {\bibinfo {author} {\bibfnamefont {J.}~\bibnamefont
  {{Grollier}}}, \bibinfo {author} {\bibfnamefont {D.}~\bibnamefont
  {{Querlioz}}}, \ and\ \bibinfo {author} {\bibfnamefont {M.~D.}\ \bibnamefont
  {{Stiles}}},\ }\href {\doibase 10.1109/JPROC.2016.2597152} {\bibfield
  {journal} {\bibinfo  {journal} {Proc. IEEE}\ }\textbf {\bibinfo {volume}
  {104}},\ \bibinfo {pages} {2024} (\bibinfo {year} {2016})}\BibitemShut
  {NoStop}%
\bibitem [{\citenamefont {Sengupta}\ and\ \citenamefont
  {Roy}(2018)}]{sengupta2018neuromorphic}%
  \BibitemOpen
  \bibfield  {author} {\bibinfo {author} {\bibfnamefont {A.}~\bibnamefont
  {Sengupta}}\ and\ \bibinfo {author} {\bibfnamefont {K.}~\bibnamefont {Roy}},\
  }\href {\doibase https://doi.org/10.7567/APEX.11.030101} {\bibfield
  {journal} {\bibinfo  {journal} {Appl. Phys. Express}\ }\textbf {\bibinfo
  {volume} {11}},\ \bibinfo {pages} {030101} (\bibinfo {year}
  {2018})}\BibitemShut {NoStop}%
\bibitem [{\citenamefont {Umesh}\ and\ \citenamefont
  {Mittal}(2019)}]{UMESH2019survey}%
  \BibitemOpen
  \bibfield  {author} {\bibinfo {author} {\bibfnamefont {S.}~\bibnamefont
  {Umesh}}\ and\ \bibinfo {author} {\bibfnamefont {S.}~\bibnamefont {Mittal}},\
  }\href {\doibase https://doi.org/10.1016/j.sysarc.2018.11.005} {\bibfield
  {journal} {\bibinfo  {journal} {J. Syst. Archit.}\ }\textbf {\bibinfo
  {volume} {97}},\ \bibinfo {pages} {349 } (\bibinfo {year}
  {2019})}\BibitemShut {NoStop}%
\bibitem [{\citenamefont {Grollier}\ \emph {et~al.}(2020)\citenamefont
  {Grollier}, \citenamefont {Querlioz}, \citenamefont {Camsari}, \citenamefont
  {Everschor-Sitte}, \citenamefont {Fukami},\ and\ \citenamefont
  {Stiles}}]{grollier2020neuromorphic}%
  \BibitemOpen
  \bibfield  {author} {\bibinfo {author} {\bibfnamefont {J.}~\bibnamefont
  {Grollier}}, \bibinfo {author} {\bibfnamefont {D.}~\bibnamefont {Querlioz}},
  \bibinfo {author} {\bibfnamefont {K.}~\bibnamefont {Camsari}}, \bibinfo
  {author} {\bibfnamefont {K.}~\bibnamefont {Everschor-Sitte}}, \bibinfo
  {author} {\bibfnamefont {S.}~\bibnamefont {Fukami}}, \ and\ \bibinfo {author}
  {\bibfnamefont {M.}~\bibnamefont {Stiles}},\ }\href
  {https://doi.org/10.1038/s41928-019-0360-9} {\bibfield  {journal} {\bibinfo
  {journal} {Nat. Electron.}\ }\textbf {\bibinfo {volume} {3}},\ \bibinfo
  {pages} {360} (\bibinfo {year} {2020})}\BibitemShut {NoStop}%
\bibitem [{\citenamefont {Kurenkov}\ \emph {et~al.}(2020)\citenamefont
  {Kurenkov}, \citenamefont {Fukami},\ and\ \citenamefont
  {Ohno}}]{KurenkovReview}%
  \BibitemOpen
  \bibfield  {author} {\bibinfo {author} {\bibfnamefont {A.}~\bibnamefont
  {Kurenkov}}, \bibinfo {author} {\bibfnamefont {S.}~\bibnamefont {Fukami}}, \
  and\ \bibinfo {author} {\bibfnamefont {H.}~\bibnamefont {Ohno}},\ }\href
  {\doibase 10.1063/5.0009482} {\bibfield  {journal} {\bibinfo  {journal} {J.
  Appl. Phys.}\ }\textbf {\bibinfo {volume} {128}},\ \bibinfo {pages} {010902}
  (\bibinfo {year} {2020})}\BibitemShut {NoStop}%
\bibitem [{\citenamefont {Allwood}\ \emph {et~al.}(2005)\citenamefont
  {Allwood}, \citenamefont {Xiong}, \citenamefont {Faulkner}, \citenamefont
  {Atkinson}, \citenamefont {Petit},\ and\ \citenamefont
  {Cowburn}}]{Allwood2005}%
  \BibitemOpen
  \bibfield  {author} {\bibinfo {author} {\bibfnamefont {D.~A.}\ \bibnamefont
  {Allwood}}, \bibinfo {author} {\bibfnamefont {G.}~\bibnamefont {Xiong}},
  \bibinfo {author} {\bibfnamefont {C.~C.}\ \bibnamefont {Faulkner}}, \bibinfo
  {author} {\bibfnamefont {D.}~\bibnamefont {Atkinson}}, \bibinfo {author}
  {\bibfnamefont {D.}~\bibnamefont {Petit}}, \ and\ \bibinfo {author}
  {\bibfnamefont {R.~P.}\ \bibnamefont {Cowburn}},\ }\href {\doibase
  10.1126/science.1108813} {\bibfield  {journal} {\bibinfo  {journal}
  {Science}\ }\textbf {\bibinfo {volume} {309}},\ \bibinfo {pages} {1688}
  (\bibinfo {year} {2005})}\BibitemShut {NoStop}%
\bibitem [{\citenamefont {Parkin}\ \emph {et~al.}(2008)\citenamefont {Parkin},
  \citenamefont {Hayashi},\ and\ \citenamefont {Thomas}}]{parkin2008DW}%
  \BibitemOpen
  \bibfield  {author} {\bibinfo {author} {\bibfnamefont {S.~S.}\ \bibnamefont
  {Parkin}}, \bibinfo {author} {\bibfnamefont {M.}~\bibnamefont {Hayashi}}, \
  and\ \bibinfo {author} {\bibfnamefont {L.}~\bibnamefont {Thomas}},\ }\href
  {https://science.sciencemag.org/content/320/5873/190} {\bibfield  {journal}
  {\bibinfo  {journal} {Science}\ }\textbf {\bibinfo {volume} {320}},\ \bibinfo
  {pages} {190} (\bibinfo {year} {2008})}\BibitemShut {NoStop}%
\bibitem [{\citenamefont {Fert}(2008)}]{Fert2008review}%
  \BibitemOpen
  \bibfield  {author} {\bibinfo {author} {\bibfnamefont {A.}~\bibnamefont
  {Fert}},\ }\href {\doibase 10.1103/RevModPhys.80.1517} {\bibfield  {journal}
  {\bibinfo  {journal} {Rev. Mod. Phys.}\ }\textbf {\bibinfo {volume} {80}},\
  \bibinfo {pages} {1517} (\bibinfo {year} {2008})}\BibitemShut {NoStop}%
\bibitem [{\citenamefont {Fert}\ \emph {et~al.}(2013)\citenamefont {Fert},
  \citenamefont {Cros},\ and\ \citenamefont {Sampaio}}]{fert2013skyrmions}%
  \BibitemOpen
  \bibfield  {author} {\bibinfo {author} {\bibfnamefont {A.}~\bibnamefont
  {Fert}}, \bibinfo {author} {\bibfnamefont {V.}~\bibnamefont {Cros}}, \ and\
  \bibinfo {author} {\bibfnamefont {J.}~\bibnamefont {Sampaio}},\ }\href
  {\doibase https://doi.org/10.1038/nnano.2013.29} {\bibfield  {journal}
  {\bibinfo  {journal} {Nat. Nanotechnol.}\ }\textbf {\bibinfo {volume} {8}},\
  \bibinfo {pages} {152} (\bibinfo {year} {2013})}\BibitemShut {NoStop}%
\bibitem [{\citenamefont {Bhatti}\ \emph {et~al.}(2017)\citenamefont {Bhatti},
  \citenamefont {Sbiaa}, \citenamefont {Hirohata}, \citenamefont {Ohno},
  \citenamefont {Fukami},\ and\ \citenamefont {Piramanayagam}}]{Bhatti2017RAM}%
  \BibitemOpen
  \bibfield  {author} {\bibinfo {author} {\bibfnamefont {S.}~\bibnamefont
  {Bhatti}}, \bibinfo {author} {\bibfnamefont {R.}~\bibnamefont {Sbiaa}},
  \bibinfo {author} {\bibfnamefont {A.}~\bibnamefont {Hirohata}}, \bibinfo
  {author} {\bibfnamefont {H.}~\bibnamefont {Ohno}}, \bibinfo {author}
  {\bibfnamefont {S.}~\bibnamefont {Fukami}}, \ and\ \bibinfo {author}
  {\bibfnamefont {S.}~\bibnamefont {Piramanayagam}},\ }\href@noop {} {\bibfield
   {journal} {\bibinfo  {journal} {Mater. Today}\ }\textbf {\bibinfo {volume}
  {20}},\ \bibinfo {pages} {530} (\bibinfo {year} {2017})}\BibitemShut
  {NoStop}%
\bibitem [{\citenamefont {Krzysteczko}\ \emph {et~al.}(2012)\citenamefont
  {Krzysteczko}, \citenamefont {Münchenberger}, \citenamefont {Schäfers},
  \citenamefont {Reiss},\ and\ \citenamefont {Thomas}}]{Krzysteczko2012MTJ}%
  \BibitemOpen
  \bibfield  {author} {\bibinfo {author} {\bibfnamefont {P.}~\bibnamefont
  {Krzysteczko}}, \bibinfo {author} {\bibfnamefont {J.}~\bibnamefont
  {Münchenberger}}, \bibinfo {author} {\bibfnamefont {M.}~\bibnamefont
  {Schäfers}}, \bibinfo {author} {\bibfnamefont {G.}~\bibnamefont {Reiss}}, \
  and\ \bibinfo {author} {\bibfnamefont {A.}~\bibnamefont {Thomas}},\ }\href
  {\doibase 10.1002/adma.201103723} {\bibfield  {journal} {\bibinfo  {journal}
  {Adv. Mater.}\ }\textbf {\bibinfo {volume} {24}},\ \bibinfo {pages} {762}
  (\bibinfo {year} {2012})}\BibitemShut {NoStop}%
\bibitem [{\citenamefont {{Sengupta}}\ and\ \citenamefont
  {{Roy}}(2015)}]{Sengupta2015MTJ}%
  \BibitemOpen
  \bibfield  {author} {\bibinfo {author} {\bibfnamefont {A.}~\bibnamefont
  {{Sengupta}}}\ and\ \bibinfo {author} {\bibfnamefont {K.}~\bibnamefont
  {{Roy}}},\ }in\ \href {\doibase 10.1109/IJCNN.2015.7280306} {\emph {\bibinfo
  {booktitle} {2015 International Joint Conference on Neural Networks (IJCNN),
  Killarney}}}\ (\bibinfo {year} {2015})\ pp.\ \bibinfo {pages}
  {1--7}\BibitemShut {NoStop}%
\bibitem [{\citenamefont {{Vincent}}\ \emph {et~al.}(2015)\citenamefont
  {{Vincent}}, \citenamefont {{Larroque}}, \citenamefont {{Locatelli}},
  \citenamefont {{Ben Romdhane}}, \citenamefont {{Bichler}}, \citenamefont
  {{Gamrat}}, \citenamefont {{Zhao}}, \citenamefont {{Klein}}, \citenamefont
  {{Galdin-Retailleau}},\ and\ \citenamefont {{Querlioz}}}]{Vincent2015MTJ}%
  \BibitemOpen
  \bibfield  {author} {\bibinfo {author} {\bibfnamefont {A.~F.}\ \bibnamefont
  {{Vincent}}}, \bibinfo {author} {\bibfnamefont {J.}~\bibnamefont
  {{Larroque}}}, \bibinfo {author} {\bibfnamefont {N.}~\bibnamefont
  {{Locatelli}}}, \bibinfo {author} {\bibfnamefont {N.}~\bibnamefont {{Ben
  Romdhane}}}, \bibinfo {author} {\bibfnamefont {O.}~\bibnamefont {{Bichler}}},
  \bibinfo {author} {\bibfnamefont {C.}~\bibnamefont {{Gamrat}}}, \bibinfo
  {author} {\bibfnamefont {W.~S.}\ \bibnamefont {{Zhao}}}, \bibinfo {author}
  {\bibfnamefont {J.}~\bibnamefont {{Klein}}}, \bibinfo {author} {\bibfnamefont
  {S.}~\bibnamefont {{Galdin-Retailleau}}}, \ and\ \bibinfo {author}
  {\bibfnamefont {D.}~\bibnamefont {{Querlioz}}},\ }\href {\doibase
  10.1109/TBCAS.2015.2414423} {\bibfield  {journal} {\bibinfo  {journal} {IEEE
  Trans. Biomed. Circuits Syst.}\ }\textbf {\bibinfo {volume} {9}},\ \bibinfo
  {pages} {166} (\bibinfo {year} {2015})}\BibitemShut {NoStop}%
\bibitem [{\citenamefont {{Zhang}}\ \emph
  {et~al.}(2016{\natexlab{a}})\citenamefont {{Zhang}}, \citenamefont {{Zeng}},
  \citenamefont {{Cao}}, \citenamefont {{Wang}}, \citenamefont {{Peng}},
  \citenamefont {{Zhang}}, \citenamefont {{Zhang}}, \citenamefont {{Klein}},
  \citenamefont {{Wang}},\ and\ \citenamefont {{Zhao}}}]{Zhang2016MTJ}%
  \BibitemOpen
  \bibfield  {author} {\bibinfo {author} {\bibfnamefont {D.}~\bibnamefont
  {{Zhang}}}, \bibinfo {author} {\bibfnamefont {L.}~\bibnamefont {{Zeng}}},
  \bibinfo {author} {\bibfnamefont {K.}~\bibnamefont {{Cao}}}, \bibinfo
  {author} {\bibfnamefont {M.}~\bibnamefont {{Wang}}}, \bibinfo {author}
  {\bibfnamefont {S.}~\bibnamefont {{Peng}}}, \bibinfo {author} {\bibfnamefont
  {Y.}~\bibnamefont {{Zhang}}}, \bibinfo {author} {\bibfnamefont
  {Y.}~\bibnamefont {{Zhang}}}, \bibinfo {author} {\bibfnamefont
  {J.}~\bibnamefont {{Klein}}}, \bibinfo {author} {\bibfnamefont
  {Y.}~\bibnamefont {{Wang}}}, \ and\ \bibinfo {author} {\bibfnamefont
  {W.}~\bibnamefont {{Zhao}}},\ }\href {\doibase 10.1109/TBCAS.2016.2533798}
  {\bibfield  {journal} {\bibinfo  {journal} {IEEE Trans. Biomed. Circuits and
  Syst.}\ }\textbf {\bibinfo {volume} {10}},\ \bibinfo {pages} {828} (\bibinfo
  {year} {2016}{\natexlab{a}})}\BibitemShut {NoStop}%
\bibitem [{\citenamefont {{Zhang}}\ \emph
  {et~al.}(2016{\natexlab{b}})\citenamefont {{Zhang}}, \citenamefont {{Zeng}},
  \citenamefont {{Zhang}}, \citenamefont {{Zhao}},\ and\ \citenamefont
  {{Klein}}}]{Zhang2016MTJ2}%
  \BibitemOpen
  \bibfield  {author} {\bibinfo {author} {\bibfnamefont {D.}~\bibnamefont
  {{Zhang}}}, \bibinfo {author} {\bibfnamefont {L.}~\bibnamefont {{Zeng}}},
  \bibinfo {author} {\bibfnamefont {Y.}~\bibnamefont {{Zhang}}}, \bibinfo
  {author} {\bibfnamefont {W.}~\bibnamefont {{Zhao}}}, \ and\ \bibinfo {author}
  {\bibfnamefont {J.~O.}\ \bibnamefont {{Klein}}},\ }in\ \href {\doibase
  10.1145/2950067.2950105} {\emph {\bibinfo {booktitle} {2016 IEEE/ACM
  International Symposium on Nanoscale Architectures (NANOARCH), Beijing}}}\
  (\bibinfo {year} {2016})\ pp.\ \bibinfo {pages} {173--178}\BibitemShut
  {NoStop}%
\bibitem [{\citenamefont {Torrejon}\ \emph {et~al.}(2017)\citenamefont
  {Torrejon}, \citenamefont {Riou}, \citenamefont {Araujo}, \citenamefont
  {Tsunegi}, \citenamefont {Khalsa}, \citenamefont {Querlioz}, \citenamefont
  {Bortolotti}, \citenamefont {Cros}, \citenamefont {Yakushiji}, \citenamefont
  {Fukushima} \emph {et~al.}}]{torrejon2017MTJ}%
  \BibitemOpen
  \bibfield  {author} {\bibinfo {author} {\bibfnamefont {J.}~\bibnamefont
  {Torrejon}}, \bibinfo {author} {\bibfnamefont {M.}~\bibnamefont {Riou}},
  \bibinfo {author} {\bibfnamefont {F.~A.}\ \bibnamefont {Araujo}}, \bibinfo
  {author} {\bibfnamefont {S.}~\bibnamefont {Tsunegi}}, \bibinfo {author}
  {\bibfnamefont {G.}~\bibnamefont {Khalsa}}, \bibinfo {author} {\bibfnamefont
  {D.}~\bibnamefont {Querlioz}}, \bibinfo {author} {\bibfnamefont
  {P.}~\bibnamefont {Bortolotti}}, \bibinfo {author} {\bibfnamefont
  {V.}~\bibnamefont {Cros}}, \bibinfo {author} {\bibfnamefont {K.}~\bibnamefont
  {Yakushiji}}, \bibinfo {author} {\bibfnamefont {A.}~\bibnamefont
  {Fukushima}},  \emph {et~al.},\ }\href {\doibase
  https://doi.org/10.1038/nature23011} {\bibfield  {journal} {\bibinfo
  {journal} {Nature (London)}\ }\textbf {\bibinfo {volume} {547}},\ \bibinfo
  {pages} {428} (\bibinfo {year} {2017})}\BibitemShut {NoStop}%
\bibitem [{\citenamefont {Kurenkov}\ \emph {et~al.}(2019)\citenamefont
  {Kurenkov}, \citenamefont {DuttaGupta}, \citenamefont {Zhang}, \citenamefont
  {Fukami}, \citenamefont {Horio},\ and\ \citenamefont {Ohno}}]{kurenkov2019}%
  \BibitemOpen
  \bibfield  {author} {\bibinfo {author} {\bibfnamefont {A.}~\bibnamefont
  {Kurenkov}}, \bibinfo {author} {\bibfnamefont {S.}~\bibnamefont
  {DuttaGupta}}, \bibinfo {author} {\bibfnamefont {C.}~\bibnamefont {Zhang}},
  \bibinfo {author} {\bibfnamefont {S.}~\bibnamefont {Fukami}}, \bibinfo
  {author} {\bibfnamefont {Y.}~\bibnamefont {Horio}}, \ and\ \bibinfo {author}
  {\bibfnamefont {H.}~\bibnamefont {Ohno}},\ }\href {\doibase
  https://doi.org/10.1002/adma.201900636} {\bibfield  {journal} {\bibinfo
  {journal} {Adv. Mater.}\ }\textbf {\bibinfo {volume} {31}},\ \bibinfo {pages}
  {1900636} (\bibinfo {year} {2019})}\BibitemShut {NoStop}%
\bibitem [{\citenamefont {Khymyn}\ \emph {et~al.}(2018)\citenamefont {Khymyn},
  \citenamefont {Lisenkov}, \citenamefont {Voorheis}, \citenamefont
  {Sulymenko}, \citenamefont {Prokopenko}, \citenamefont {Tiberkevich},
  \citenamefont {Akerman},\ and\ \citenamefont {Slavin}}]{Khymyn2018ST}%
  \BibitemOpen
  \bibfield  {author} {\bibinfo {author} {\bibfnamefont {R.}~\bibnamefont
  {Khymyn}}, \bibinfo {author} {\bibfnamefont {I.}~\bibnamefont {Lisenkov}},
  \bibinfo {author} {\bibfnamefont {J.}~\bibnamefont {Voorheis}}, \bibinfo
  {author} {\bibfnamefont {O.}~\bibnamefont {Sulymenko}}, \bibinfo {author}
  {\bibfnamefont {O.}~\bibnamefont {Prokopenko}}, \bibinfo {author}
  {\bibfnamefont {V.}~\bibnamefont {Tiberkevich}}, \bibinfo {author}
  {\bibfnamefont {J.}~\bibnamefont {Akerman}}, \ and\ \bibinfo {author}
  {\bibfnamefont {A.}~\bibnamefont {Slavin}},\ }\href {\doibase
  https://doi.org/10.1038/s41598-018-33697-0} {\bibfield  {journal} {\bibinfo
  {journal} {Sci. Rep.}\ }\textbf {\bibinfo {volume} {8}},\ \bibinfo {pages}
  {15727} (\bibinfo {year} {2018})}\BibitemShut {NoStop}%
\bibitem [{\citenamefont {{Sulymenko}}\ and\ \citenamefont
  {{Prokopenko}}(2019)}]{Sulymenko2019Logic}%
  \BibitemOpen
  \bibfield  {author} {\bibinfo {author} {\bibfnamefont {O.~R.}\ \bibnamefont
  {{Sulymenko}}}\ and\ \bibinfo {author} {\bibfnamefont {O.~V.}\ \bibnamefont
  {{Prokopenko}}},\ }in\ \href {\doibase 10.1109/ELNANO.2019.8783266} {\emph
  {\bibinfo {booktitle} {2019 IEEE 39th International Conference on Electronics
  and Nanotechnology (ELNANO), Kyiv, Ukraine}}}\ (\bibinfo {year} {2019})\ pp.\
  \bibinfo {pages} {132--137}\BibitemShut {NoStop}%
\bibitem [{\citenamefont {Fukami}\ and\ \citenamefont
  {Ohno}(2018)}]{Fukami2018ST}%
  \BibitemOpen
  \bibfield  {author} {\bibinfo {author} {\bibfnamefont {S.}~\bibnamefont
  {Fukami}}\ and\ \bibinfo {author} {\bibfnamefont {H.}~\bibnamefont {Ohno}},\
  }\href {\doibase 10.1063/1.5042317} {\bibfield  {journal} {\bibinfo
  {journal} {J. Appl. Phys.}\ }\textbf {\bibinfo {volume} {124}},\ \bibinfo
  {pages} {151904} (\bibinfo {year} {2018})}\BibitemShut {NoStop}%
\bibitem [{\citenamefont {Yue}\ \emph {et~al.}(2019)\citenamefont {Yue},
  \citenamefont {Liu}, \citenamefont {Lake},\ and\ \citenamefont
  {Parker}}]{Yue2019DW}%
  \BibitemOpen
  \bibfield  {author} {\bibinfo {author} {\bibfnamefont {K.}~\bibnamefont
  {Yue}}, \bibinfo {author} {\bibfnamefont {Y.}~\bibnamefont {Liu}}, \bibinfo
  {author} {\bibfnamefont {R.~K.}\ \bibnamefont {Lake}}, \ and\ \bibinfo
  {author} {\bibfnamefont {A.~C.}\ \bibnamefont {Parker}},\ }\href
  {https://advances.sciencemag.org/content/5/4/eaau8170} {\bibfield  {journal}
  {\bibinfo  {journal} {Sci. Adv.}\ }\textbf {\bibinfo {volume} {5}},\ \bibinfo
  {pages} {eaau8170} (\bibinfo {year} {2019})}\BibitemShut {NoStop}%
\bibitem [{\citenamefont {Sharad}\ \emph {et~al.}(2013)\citenamefont {Sharad},
  \citenamefont {Fan},\ and\ \citenamefont {Roy}}]{Sharad2013DW}%
  \BibitemOpen
  \bibfield  {author} {\bibinfo {author} {\bibfnamefont {M.}~\bibnamefont
  {Sharad}}, \bibinfo {author} {\bibfnamefont {D.}~\bibnamefont {Fan}}, \ and\
  \bibinfo {author} {\bibfnamefont {K.}~\bibnamefont {Roy}},\ }\href {\doibase
  10.1063/1.4838096} {\bibfield  {journal} {\bibinfo  {journal} {J. Appl.
  Phys.}\ }\textbf {\bibinfo {volume} {114}},\ \bibinfo {pages} {234906}
  (\bibinfo {year} {2013})}\BibitemShut {NoStop}%
\bibitem [{\citenamefont {{Cui}}\ \emph {et~al.}(2019)\citenamefont {{Cui}},
  \citenamefont {{Akinola}}, \citenamefont {{Hassan}}, \citenamefont
  {{Bennett}}, \citenamefont {{Marinella}}, \citenamefont {{Friedman}},\ and\
  \citenamefont {{Incorvia}}}]{Cui2019}%
  \BibitemOpen
  \bibfield  {author} {\bibinfo {author} {\bibfnamefont {C.}~\bibnamefont
  {{Cui}}}, \bibinfo {author} {\bibfnamefont {O.~G.}\ \bibnamefont
  {{Akinola}}}, \bibinfo {author} {\bibfnamefont {N.}~\bibnamefont {{Hassan}}},
  \bibinfo {author} {\bibfnamefont {C.~H.}\ \bibnamefont {{Bennett}}}, \bibinfo
  {author} {\bibfnamefont {M.~J.}\ \bibnamefont {{Marinella}}}, \bibinfo
  {author} {\bibfnamefont {J.~S.}\ \bibnamefont {{Friedman}}}, \ and\ \bibinfo
  {author} {\bibfnamefont {J.~A.~C.}\ \bibnamefont {{Incorvia}}},\ }\href@noop
  {} {\  (\bibinfo {year} {2019})},\ \Eprint {http://arxiv.org/abs/1912.04505}
  {arXiv:1912.04505} \BibitemShut {NoStop}%
\bibitem [{\citenamefont {Huang}\ \emph {et~al.}(2017)\citenamefont {Huang},
  \citenamefont {Kang}, \citenamefont {Zhang}, \citenamefont {Zhou},\ and\
  \citenamefont {Zhao}}]{Huang2017Skyr}%
  \BibitemOpen
  \bibfield  {author} {\bibinfo {author} {\bibfnamefont {Y.}~\bibnamefont
  {Huang}}, \bibinfo {author} {\bibfnamefont {W.}~\bibnamefont {Kang}},
  \bibinfo {author} {\bibfnamefont {X.}~\bibnamefont {Zhang}}, \bibinfo
  {author} {\bibfnamefont {Y.}~\bibnamefont {Zhou}}, \ and\ \bibinfo {author}
  {\bibfnamefont {W.}~\bibnamefont {Zhao}},\ }\href {\doibase
  10.1088/1361-6528/aa5838} {\bibfield  {journal} {\bibinfo  {journal}
  {Nanotechnology}\ }\textbf {\bibinfo {volume} {28}},\ \bibinfo {pages}
  {08LT02} (\bibinfo {year} {2017})}\BibitemShut {NoStop}%
\bibitem [{\citenamefont {Li}\ \emph {et~al.}(2017)\citenamefont {Li},
  \citenamefont {Kang}, \citenamefont {Huang}, \citenamefont {Zhang},
  \citenamefont {Zhou},\ and\ \citenamefont {Zhao}}]{Li2017Skyr}%
  \BibitemOpen
  \bibfield  {author} {\bibinfo {author} {\bibfnamefont {S.}~\bibnamefont
  {Li}}, \bibinfo {author} {\bibfnamefont {W.}~\bibnamefont {Kang}}, \bibinfo
  {author} {\bibfnamefont {Y.}~\bibnamefont {Huang}}, \bibinfo {author}
  {\bibfnamefont {X.}~\bibnamefont {Zhang}}, \bibinfo {author} {\bibfnamefont
  {Y.}~\bibnamefont {Zhou}}, \ and\ \bibinfo {author} {\bibfnamefont
  {W.}~\bibnamefont {Zhao}},\ }\href {\doibase 10.1088/1361-6528/aa7af5}
  {\bibfield  {journal} {\bibinfo  {journal} {Nanotechnology}\ }\textbf
  {\bibinfo {volume} {28}},\ \bibinfo {pages} {31LT01} (\bibinfo {year}
  {2017})}\BibitemShut {NoStop}%
\bibitem [{\citenamefont {{Chen}}\ \emph {et~al.}(2018)\citenamefont {{Chen}},
  \citenamefont {{Sengupta}},\ and\ \citenamefont {{Roy}}}]{Chen2018Skyr}%
  \BibitemOpen
  \bibfield  {author} {\bibinfo {author} {\bibfnamefont {M.}~\bibnamefont
  {{Chen}}}, \bibinfo {author} {\bibfnamefont {A.}~\bibnamefont {{Sengupta}}},
  \ and\ \bibinfo {author} {\bibfnamefont {K.}~\bibnamefont {{Roy}}},\ }\href
  {\doibase 10.1109/TMAG.2018.2845890} {\bibfield  {journal} {\bibinfo
  {journal} {IEEE Trans. Magn.}\ }\textbf {\bibinfo {volume} {54}},\ \bibinfo
  {pages} {1} (\bibinfo {year} {2018})}\BibitemShut {NoStop}%
\bibitem [{\citenamefont {{Zeng}}\ \emph {et~al.}(2016)\citenamefont {{Zeng}},
  \citenamefont {{Zhang}}, \citenamefont {{Zhang}}, \citenamefont {{Gong}},
  \citenamefont {{Gao}}, \citenamefont {{Tu}}, \citenamefont {{Yu}},\ and\
  \citenamefont {{Zhao}}}]{Zeng2016SW}%
  \BibitemOpen
  \bibfield  {author} {\bibinfo {author} {\bibfnamefont {L.}~\bibnamefont
  {{Zeng}}}, \bibinfo {author} {\bibfnamefont {D.}~\bibnamefont {{Zhang}}},
  \bibinfo {author} {\bibfnamefont {Y.}~\bibnamefont {{Zhang}}}, \bibinfo
  {author} {\bibfnamefont {F.}~\bibnamefont {{Gong}}}, \bibinfo {author}
  {\bibfnamefont {T.}~\bibnamefont {{Gao}}}, \bibinfo {author} {\bibfnamefont
  {S.}~\bibnamefont {{Tu}}}, \bibinfo {author} {\bibfnamefont {H.}~\bibnamefont
  {{Yu}}}, \ and\ \bibinfo {author} {\bibfnamefont {W.}~\bibnamefont
  {{Zhao}}},\ }in\ \href {\doibase 10.1109/ISCAS.2016.7527391} {\emph {\bibinfo
  {booktitle} {2016 IEEE International Symposium on Circuits and Systems
  (ISCAS), Montreal, QC}}}\ (\bibinfo {year} {2016})\ pp.\ \bibinfo {pages}
  {918--921}\BibitemShut {NoStop}%
\bibitem [{\citenamefont {{Katayama}}\ \emph {et~al.}(2016)\citenamefont
  {{Katayama}}, \citenamefont {{Yamane}}, \citenamefont {{Nakano}},
  \citenamefont {{Nakane}},\ and\ \citenamefont {{Tanaka}}}]{Katayama2016SW}%
  \BibitemOpen
  \bibfield  {author} {\bibinfo {author} {\bibfnamefont {Y.}~\bibnamefont
  {{Katayama}}}, \bibinfo {author} {\bibfnamefont {T.}~\bibnamefont
  {{Yamane}}}, \bibinfo {author} {\bibfnamefont {D.}~\bibnamefont {{Nakano}}},
  \bibinfo {author} {\bibfnamefont {R.}~\bibnamefont {{Nakane}}}, \ and\
  \bibinfo {author} {\bibfnamefont {G.}~\bibnamefont {{Tanaka}}},\ }\href
  {\doibase 10.1109/TNANO.2016.2545690} {\bibfield  {journal} {\bibinfo
  {journal} {IEEE Trans. Nanotechnol.}\ }\textbf {\bibinfo {volume} {15}},\
  \bibinfo {pages} {762} (\bibinfo {year} {2016})}\BibitemShut {NoStop}%
\bibitem [{\citenamefont {Coleman}(1985)}]{TopoConservation}%
  \BibitemOpen
  \bibfield  {author} {\bibinfo {author} {\bibfnamefont {S.}~\bibnamefont
  {Coleman}},\ }\href@noop {} {\emph {\bibinfo {title} {Aspects of symmetry:
  Selected {Erice Lectures}}}}\ (\bibinfo  {publisher} {Cambridge University
  Press},\ \bibinfo {address} {New York, NY},\ \bibinfo {year}
  {1985})\BibitemShut {NoStop}%
\bibitem [{\citenamefont {Zee}(2010)}]{NutshellQFT}%
  \BibitemOpen
  \bibfield  {author} {\bibinfo {author} {\bibfnamefont {A.}~\bibnamefont
  {Zee}},\ }\href@noop {} {\emph {\bibinfo {title} {Quantum Field Theory in a
  Nutshell, 2nd edition}}}\ (\bibinfo  {publisher} {Princeton University
  Press},\ \bibinfo {address} {Princeton, NJ},\ \bibinfo {year}
  {2010})\BibitemShut {NoStop}%
\bibitem [{\citenamefont {Baltz}\ \emph {et~al.}(2018)\citenamefont {Baltz},
  \citenamefont {Manchon}, \citenamefont {Tsoi}, \citenamefont {Moriyama},
  \citenamefont {Ono},\ and\ \citenamefont {Tserkovnyak}}]{Baltz2018AFreview}%
  \BibitemOpen
  \bibfield  {author} {\bibinfo {author} {\bibfnamefont {V.}~\bibnamefont
  {Baltz}}, \bibinfo {author} {\bibfnamefont {A.}~\bibnamefont {Manchon}},
  \bibinfo {author} {\bibfnamefont {M.}~\bibnamefont {Tsoi}}, \bibinfo {author}
  {\bibfnamefont {T.}~\bibnamefont {Moriyama}}, \bibinfo {author}
  {\bibfnamefont {T.}~\bibnamefont {Ono}}, \ and\ \bibinfo {author}
  {\bibfnamefont {Y.}~\bibnamefont {Tserkovnyak}},\ }\href {\doibase
  10.1103/RevModPhys.90.015005} {\bibfield  {journal} {\bibinfo  {journal}
  {Rev. Mod. Phys.}\ }\textbf {\bibinfo {volume} {90}},\ \bibinfo {pages}
  {015005} (\bibinfo {year} {2018})}\BibitemShut {NoStop}%
\bibitem [{\citenamefont {{Gilbert}}(2004)}]{Gilbert2004}%
  \BibitemOpen
  \bibfield  {author} {\bibinfo {author} {\bibfnamefont {T.~L.}\ \bibnamefont
  {{Gilbert}}},\ }\href {\doibase 10.1109/TMAG.2004.836740} {\bibfield
  {journal} {\bibinfo  {journal} {IEEE Trans. Magn.}\ }\textbf {\bibinfo
  {volume} {40}},\ \bibinfo {pages} {3443} (\bibinfo {year}
  {2004})}\BibitemShut {NoStop}%
\bibitem [{\citenamefont {Tserkovnyak}(2018)}]{Tserkovnyak2018Perspective}%
  \BibitemOpen
  \bibfield  {author} {\bibinfo {author} {\bibfnamefont {Y.}~\bibnamefont
  {Tserkovnyak}},\ }\href {\doibase 10.1063/1.5054123} {\bibfield  {journal}
  {\bibinfo  {journal} {J. Appl. Phys.}\ }\textbf {\bibinfo {volume} {124}},\
  \bibinfo {pages} {190901} (\bibinfo {year} {2018})}\BibitemShut {NoStop}%
\bibitem [{\citenamefont {Hill}\ \emph {et~al.}(2018)\citenamefont {Hill},
  \citenamefont {Kim},\ and\ \citenamefont {Tserkovnyak}}]{Hill2018LJJ}%
  \BibitemOpen
  \bibfield  {author} {\bibinfo {author} {\bibfnamefont {D.}~\bibnamefont
  {Hill}}, \bibinfo {author} {\bibfnamefont {S.~K.}\ \bibnamefont {Kim}}, \
  and\ \bibinfo {author} {\bibfnamefont {Y.}~\bibnamefont {Tserkovnyak}},\
  }\href {\doibase 10.1103/PhysRevLett.121.037202} {\bibfield  {journal}
  {\bibinfo  {journal} {Phys. Rev. Lett.}\ }\textbf {\bibinfo {volume} {121}},\
  \bibinfo {pages} {037202} (\bibinfo {year} {2018})}\BibitemShut {NoStop}%
\bibitem [{\citenamefont {Khymyn}\ \emph {et~al.}(2017)\citenamefont {Khymyn},
  \citenamefont {Lisenkov}, \citenamefont {Tiberkevich}, \citenamefont
  {Ivanov},\ and\ \citenamefont {Slavin}}]{khymyn2017}%
  \BibitemOpen
  \bibfield  {author} {\bibinfo {author} {\bibfnamefont {R.}~\bibnamefont
  {Khymyn}}, \bibinfo {author} {\bibfnamefont {I.}~\bibnamefont {Lisenkov}},
  \bibinfo {author} {\bibfnamefont {V.}~\bibnamefont {Tiberkevich}}, \bibinfo
  {author} {\bibfnamefont {B.~A.}\ \bibnamefont {Ivanov}}, \ and\ \bibinfo
  {author} {\bibfnamefont {A.}~\bibnamefont {Slavin}},\ }\href
  {https://www.nature.com/articles/srep43705} {\bibfield  {journal} {\bibinfo
  {journal} {Sci. Rep.}\ }\textbf {\bibinfo {volume} {7}},\ \bibinfo {pages}
  {43705} (\bibinfo {year} {2017})}\BibitemShut {NoStop}%
\bibitem [{\citenamefont {Haldane}(1983)}]{Haldane1983Lorentz}%
  \BibitemOpen
  \bibfield  {author} {\bibinfo {author} {\bibfnamefont {F.~D.~M.}\
  \bibnamefont {Haldane}},\ }\href {\doibase 10.1103/PhysRevLett.50.1153}
  {\bibfield  {journal} {\bibinfo  {journal} {Phys. Rev. Lett.}\ }\textbf
  {\bibinfo {volume} {50}},\ \bibinfo {pages} {1153} (\bibinfo {year}
  {1983})}\BibitemShut {NoStop}%
\bibitem [{\citenamefont {Bar'yakhtar}\ \emph {et~al.}(1980)\citenamefont
  {Bar'yakhtar}, \citenamefont {Ivanov},\ and\ \citenamefont
  {Sukstanskii}}]{Nonlinear1980}%
  \BibitemOpen
  \bibfield  {author} {\bibinfo {author} {\bibfnamefont {V.~G.}\ \bibnamefont
  {Bar'yakhtar}}, \bibinfo {author} {\bibfnamefont {B.~A.}\ \bibnamefont
  {Ivanov}}, \ and\ \bibinfo {author} {\bibfnamefont {A.~L.}\ \bibnamefont
  {Sukstanskii}},\ }\href {http://www.jetp.ac.ru/cgi-bin/dn/e_051_04_0757.pdf}
  {\bibfield  {journal} {\bibinfo  {journal} {J. Exp. Theor. Phys.}\ }\textbf
  {\bibinfo {volume} {51}},\ \bibinfo {pages} {757} (\bibinfo {year}
  {1980})}\BibitemShut {NoStop}%
\bibitem [{\citenamefont {Hodgkin}\ and\ \citenamefont
  {Huxley}(1952)}]{hodgkin1952propagation}%
  \BibitemOpen
  \bibfield  {author} {\bibinfo {author} {\bibfnamefont {A.~L.}\ \bibnamefont
  {Hodgkin}}\ and\ \bibinfo {author} {\bibfnamefont {A.~F.}\ \bibnamefont
  {Huxley}},\ }\href {\doibase https://doi.org/10.1098/rspb.1952.0054}
  {\bibfield  {journal} {\bibinfo  {journal} {Proc. R. Soc. B}\ }\textbf
  {\bibinfo {volume} {140}},\ \bibinfo {pages} {177} (\bibinfo {year}
  {1952})}\BibitemShut {NoStop}%
\bibitem [{\citenamefont {Kim}\ \emph {et~al.}(2016)\citenamefont {Kim},
  \citenamefont {Takei},\ and\ \citenamefont {Tserkovnyak}}]{Kim2016PhaseSlip}%
  \BibitemOpen
  \bibfield  {author} {\bibinfo {author} {\bibfnamefont {S.~K.}\ \bibnamefont
  {Kim}}, \bibinfo {author} {\bibfnamefont {S.}~\bibnamefont {Takei}}, \ and\
  \bibinfo {author} {\bibfnamefont {Y.}~\bibnamefont {Tserkovnyak}},\ }\href
  {\doibase 10.1103/PhysRevB.93.020402} {\bibfield  {journal} {\bibinfo
  {journal} {Phys. Rev. B}\ }\textbf {\bibinfo {volume} {93}},\ \bibinfo
  {pages} {020402(R)} (\bibinfo {year} {2016})}\BibitemShut {NoStop}%
\bibitem [{\citenamefont {Ivanov}\ \emph {et~al.}(1998)\citenamefont {Ivanov},
  \citenamefont {Kolezhuk},\ and\ \citenamefont {Kireev}}]{Ivanov1998}%
  \BibitemOpen
  \bibfield  {author} {\bibinfo {author} {\bibfnamefont {B.~A.}\ \bibnamefont
  {Ivanov}}, \bibinfo {author} {\bibfnamefont {A.~K.}\ \bibnamefont
  {Kolezhuk}}, \ and\ \bibinfo {author} {\bibfnamefont {V.~E.}\ \bibnamefont
  {Kireev}},\ }\href {\doibase 10.1103/PhysRevB.58.11514} {\bibfield  {journal}
  {\bibinfo  {journal} {Phys. Rev. B}\ }\textbf {\bibinfo {volume} {58}},\
  \bibinfo {pages} {11514} (\bibinfo {year} {1998})}\BibitemShut {NoStop}%
\bibitem [{\citenamefont {Sonin}(2010)}]{Sonin2010Review}%
  \BibitemOpen
  \bibfield  {author} {\bibinfo {author} {\bibfnamefont {E.}~\bibnamefont
  {Sonin}},\ }\href {\doibase 10.1080/00018731003739943} {\bibfield  {journal}
  {\bibinfo  {journal} {Adv. Phys.}\ }\textbf {\bibinfo {volume} {59}},\
  \bibinfo {pages} {181} (\bibinfo {year} {2010})}\BibitemShut {NoStop}%
\bibitem [{\citenamefont {Tserkovnyak}\ \emph {et~al.}(2003)\citenamefont
  {Tserkovnyak}, \citenamefont {Brataas},\ and\ \citenamefont
  {Bauer}}]{Tserkovnyak2003SpinPump}%
  \BibitemOpen
  \bibfield  {author} {\bibinfo {author} {\bibfnamefont {Y.}~\bibnamefont
  {Tserkovnyak}}, \bibinfo {author} {\bibfnamefont {A.}~\bibnamefont
  {Brataas}}, \ and\ \bibinfo {author} {\bibfnamefont {G.~E.~W.}\ \bibnamefont
  {Bauer}},\ }\href {\doibase 10.1103/PhysRevB.67.140404} {\bibfield  {journal}
  {\bibinfo  {journal} {Phys. Rev. B}\ }\textbf {\bibinfo {volume} {67}},\
  \bibinfo {pages} {140404(R)} (\bibinfo {year} {2003})}\BibitemShut {NoStop}%
\bibitem [{sup()}]{supmat}%
  \BibitemOpen
  \href@noop {} {}\bibinfo {note} {Supplementary Material, which includes Ref.
  [61, 62]}\BibitemShut {NoStop}%
\bibitem [{\citenamefont {Heinrich}\ \emph {et~al.}(2003)\citenamefont
  {Heinrich}, \citenamefont {Tserkovnyak}, \citenamefont {Woltersdorf},
  \citenamefont {Brataas}, \citenamefont {Urban},\ and\ \citenamefont
  {Bauer}}]{Heinrich2003SpinPump}%
  \BibitemOpen
  \bibfield  {author} {\bibinfo {author} {\bibfnamefont {B.}~\bibnamefont
  {Heinrich}}, \bibinfo {author} {\bibfnamefont {Y.}~\bibnamefont
  {Tserkovnyak}}, \bibinfo {author} {\bibfnamefont {G.}~\bibnamefont
  {Woltersdorf}}, \bibinfo {author} {\bibfnamefont {A.}~\bibnamefont
  {Brataas}}, \bibinfo {author} {\bibfnamefont {R.}~\bibnamefont {Urban}}, \
  and\ \bibinfo {author} {\bibfnamefont {G.~E.~W.}\ \bibnamefont {Bauer}},\
  }\href {\doibase 10.1103/PhysRevLett.90.187601} {\bibfield  {journal}
  {\bibinfo  {journal} {Phys. Rev. Lett.}\ }\textbf {\bibinfo {volume} {90}},\
  \bibinfo {pages} {187601} (\bibinfo {year} {2003})}\BibitemShut {NoStop}%
\bibitem [{\citenamefont {Izhikevich}(2000)}]{IZHIKEVICH2000bursting}%
  \BibitemOpen
  \bibfield  {author} {\bibinfo {author} {\bibfnamefont {E.~M.}\ \bibnamefont
  {Izhikevich}},\ }\href {\doibase 10.1142/S0218127400000840} {\bibfield
  {journal} {\bibinfo  {journal} {Int. J. Bifurcation and Chaos}\ }\textbf
  {\bibinfo {volume} {10}},\ \bibinfo {pages} {1171} (\bibinfo {year}
  {2000})}\BibitemShut {NoStop}%
\bibitem [{\citenamefont {Tserkovnyak}\ and\ \citenamefont
  {Xiao}(2018)}]{Tserkovnyak2018Battery}%
  \BibitemOpen
  \bibfield  {author} {\bibinfo {author} {\bibfnamefont {Y.}~\bibnamefont
  {Tserkovnyak}}\ and\ \bibinfo {author} {\bibfnamefont {J.}~\bibnamefont
  {Xiao}},\ }\href {\doibase 10.1103/PhysRevLett.121.127701} {\bibfield
  {journal} {\bibinfo  {journal} {Phys. Rev. Lett.}\ }\textbf {\bibinfo
  {volume} {121}},\ \bibinfo {pages} {127701} (\bibinfo {year}
  {2018})}\BibitemShut {NoStop}%
\bibitem [{\citenamefont {Tserkovnyak}\ \emph {et~al.}(2005)\citenamefont
  {Tserkovnyak}, \citenamefont {Brataas}, \citenamefont {Bauer},\ and\
  \citenamefont {Halperin}}]{Tserkovnyak2005Review}%
  \BibitemOpen
  \bibfield  {author} {\bibinfo {author} {\bibfnamefont {Y.}~\bibnamefont
  {Tserkovnyak}}, \bibinfo {author} {\bibfnamefont {A.}~\bibnamefont
  {Brataas}}, \bibinfo {author} {\bibfnamefont {G.~E.~W.}\ \bibnamefont
  {Bauer}}, \ and\ \bibinfo {author} {\bibfnamefont {B.~I.}\ \bibnamefont
  {Halperin}},\ }\href {\doibase 10.1103/RevModPhys.77.1375} {\bibfield
  {journal} {\bibinfo  {journal} {Rev. Mod. Phys.}\ }\textbf {\bibinfo {volume}
  {77}},\ \bibinfo {pages} {1375} (\bibinfo {year} {2005})}\BibitemShut
  {NoStop}%
\bibitem [{\citenamefont {Yuan}\ \emph {et~al.}(2018)\citenamefont {Yuan},
  \citenamefont {Zhu}, \citenamefont {Su}, \citenamefont {Yao}, \citenamefont
  {Xing}, \citenamefont {Chen}, \citenamefont {Ma}, \citenamefont {Lin},
  \citenamefont {Shi}, \citenamefont {Shindou}, \citenamefont {Xie},\ and\
  \citenamefont {Han}}]{superfluidCrO}%
  \BibitemOpen
  \bibfield  {author} {\bibinfo {author} {\bibfnamefont {W.}~\bibnamefont
  {Yuan}}, \bibinfo {author} {\bibfnamefont {Q.}~\bibnamefont {Zhu}}, \bibinfo
  {author} {\bibfnamefont {T.}~\bibnamefont {Su}}, \bibinfo {author}
  {\bibfnamefont {Y.}~\bibnamefont {Yao}}, \bibinfo {author} {\bibfnamefont
  {W.}~\bibnamefont {Xing}}, \bibinfo {author} {\bibfnamefont {Y.}~\bibnamefont
  {Chen}}, \bibinfo {author} {\bibfnamefont {Y.}~\bibnamefont {Ma}}, \bibinfo
  {author} {\bibfnamefont {X.}~\bibnamefont {Lin}}, \bibinfo {author}
  {\bibfnamefont {J.}~\bibnamefont {Shi}}, \bibinfo {author} {\bibfnamefont
  {R.}~\bibnamefont {Shindou}}, \bibinfo {author} {\bibfnamefont {X.~C.}\
  \bibnamefont {Xie}}, \ and\ \bibinfo {author} {\bibfnamefont
  {W.}~\bibnamefont {Han}},\ }\href
  {https://advances.sciencemag.org/content/4/4/eaat1098} {\bibfield  {journal}
  {\bibinfo  {journal} {Sci. Adv.}\ }\textbf {\bibinfo {volume} {4}} (\bibinfo
  {year} {2018})}\BibitemShut {NoStop}%
\bibitem [{\citenamefont {Takei}\ \emph {et~al.}(2014)\citenamefont {Takei},
  \citenamefont {Halperin}, \citenamefont {Yacoby},\ and\ \citenamefont
  {Tserkovnyak}}]{Takei2014AFM}%
  \BibitemOpen
  \bibfield  {author} {\bibinfo {author} {\bibfnamefont {S.}~\bibnamefont
  {Takei}}, \bibinfo {author} {\bibfnamefont {B.~I.}\ \bibnamefont {Halperin}},
  \bibinfo {author} {\bibfnamefont {A.}~\bibnamefont {Yacoby}}, \ and\ \bibinfo
  {author} {\bibfnamefont {Y.}~\bibnamefont {Tserkovnyak}},\ }\href {\doibase
  10.1103/PhysRevB.90.094408} {\bibfield  {journal} {\bibinfo  {journal} {Phys.
  Rev. B}\ }\textbf {\bibinfo {volume} {90}},\ \bibinfo {pages} {094408}
  (\bibinfo {year} {2014})}\BibitemShut {NoStop}%
\bibitem [{\citenamefont {Burch}\ \emph {et~al.}(2018)\citenamefont {Burch},
  \citenamefont {Mandrus},\ and\ \citenamefont {Park}}]{vanderWaals}%
  \BibitemOpen
  \bibfield  {author} {\bibinfo {author} {\bibfnamefont {K.~S.}\ \bibnamefont
  {Burch}}, \bibinfo {author} {\bibfnamefont {D.}~\bibnamefont {Mandrus}}, \
  and\ \bibinfo {author} {\bibfnamefont {J.-G.}\ \bibnamefont {Park}},\ }\href
  {https://www.nature.com/articles/s41586-018-0631-z} {\bibfield  {journal}
  {\bibinfo  {journal} {Nature (London)}\ }\textbf {\bibinfo {volume} {563}},\
  \bibinfo {pages} {47} (\bibinfo {year} {2018})}\BibitemShut {NoStop}%
\bibitem [{\citenamefont {Yang}\ \emph {et~al.}(2015)\citenamefont {Yang},
  \citenamefont {Ryu},\ and\ \citenamefont {Parkin}}]{syntheticAFM}%
  \BibitemOpen
  \bibfield  {author} {\bibinfo {author} {\bibfnamefont {S.-H.}\ \bibnamefont
  {Yang}}, \bibinfo {author} {\bibfnamefont {K.-S.}\ \bibnamefont {Ryu}}, \
  and\ \bibinfo {author} {\bibfnamefont {S.}~\bibnamefont {Parkin}},\ }\href
  {https://doi.org/10.1038/nnano.2014.324} {\bibfield  {journal} {\bibinfo
  {journal} {Nat. Nanotechnol.}\ }\textbf {\bibinfo {volume} {10}},\ \bibinfo
  {pages} {221} (\bibinfo {year} {2015})}\BibitemShut {NoStop}%
\bibitem [{\citenamefont {Tserkovnyak}\ and\ \citenamefont
  {Ochoa}(2017)}]{Tserkovnyak2017Boundary}%
  \BibitemOpen
  \bibfield  {author} {\bibinfo {author} {\bibfnamefont {Y.}~\bibnamefont
  {Tserkovnyak}}\ and\ \bibinfo {author} {\bibfnamefont {H.}~\bibnamefont
  {Ochoa}},\ }\href {\doibase 10.1103/PhysRevB.96.100402} {\bibfield  {journal}
  {\bibinfo  {journal} {Phys. Rev. B}\ }\textbf {\bibinfo {volume} {96}},\
  \bibinfo {pages} {100402(R)} (\bibinfo {year} {2017})}\BibitemShut {NoStop}%
\bibitem [{\citenamefont {Akhiezer}\ and\ \citenamefont
  {Silver}(1990)}]{Elliptic}%
  \BibitemOpen
  \bibfield  {author} {\bibinfo {author} {\bibfnamefont {N.~I.}\ \bibnamefont
  {Akhiezer}}\ and\ \bibinfo {author} {\bibfnamefont {B.}~\bibnamefont
  {Silver}},\ }\href@noop {} {\emph {\bibinfo {title} {Elements of the Theory
  of Elliptic Functions}}}\ (\bibinfo  {publisher} {American Mathematical
  Society},\ \bibinfo {address} {Providence, RI},\ \bibinfo {year}
  {1990})\BibitemShut {NoStop}%
\end{thebibliography}%

\clearpage
\onecolumngrid

\begin{center}
  \large{\textbf{Supplemental Material}}  
\end{center}
\appendix
\setcounter{equation}{0}
\renewcommand{\theequation}{A.\arabic{equation}}

\section{Simplified spin-valve dynamics}

A hierarchy of timescales is assumed for our simplified spin-valve (dendrite-metal spacer-neuron) structure. Taking the domain-wall width $\lambda \!\sim\! 10$~nm and the domain-wall velocity $v \!\sim\! 0.01u$--$0.1u$, $u \!\sim\! 10 $~km/s in the dendrite, the pumping of the spin current is on a timescale of
$t_s \!\sim\! \lambda/v \!\sim\! 0.01$--$0.1$~ns. The electron transport time across the normal-metal spacer with thickness $\ell$, i.e., the Thouless time, can be esitmated by $t_d \!\sim\! \ell^2/D \!\sim\! 0.1$~ps, where $D$ is the diffusion constant. The reaction time of the  neuron is
$t_0 \!\sim\! 1$~ns, as estimated by the harmonic-oscillator approximation in the main text. One finds
$t_d \!<\! t_s \!<\! t_0$, the angular-momentum transfer is instantaneous.

The transmission and reflection of the spin current in the metal spacer mentioned in the main text is an over-simplified description of the following dynamics.

In response to the spin dynamics at the dendrite-metal interface, the metal spacer develops an electronic spin accumulation $\mu_s$ oriented out of the easy plane of the magnet. The torques acting on the two interfaces of the spin valve is thus~\cite{Tserkovnyak2017Boundary}
\begin{equation}
\tau_1  \!=\!  \frac{\hbar g}{4\pi} (\frac{\mu_s}{\hbar} - \partial_t \phi), \text{ and }
\tau_2  \!=\!  \frac{\hbar g}{4\pi} \frac{\mu_s}{\hbar}.
\end{equation}
The torque $\tau_1$ at the dendrite-metal interface has two terms: the spin torque due to the spin accumulation, and the spin pumping, which is precisely the spin current $I_s$ in the main text. In the torque $\tau_2$ at the neuron-metal interface, the spin pumping is absent because of the slow reaction of the magnetization in the neuron. Based on the abovementioned hierarchy of timescales, our simplified situation can be captured by $- \tau_1 = \tau_2$ (an effective $\mu_s = \hbar \partial_t \phi/2$), and an effective spin-mixing conductance $g$. Unequal partitioning of the spin injection, depending on the relative interfacial impedances, as well as spin losses in the metal spacer would merely modify the effective mixing conductance g in this relation~\cite{Tserkovnyak2003SpinPump}.


\section{Exact solutions to the static sine-Gordon equation}

A static texture in equilibrium is exactly solvable by minimizing the free energy
\begin{equation}
    F \!=\! \int_0^L dx \left[ \frac{\mathcal{A}}{2} (\partial_x \phi)^2 + \frac{\mathcal{K}}{2} \sin^2 \phi \right]
\end{equation}
at low temperatures (ignoring geometric effects and entropic contributions). The solution reads~\cite{Elliptic}
\begin{equation}
    \phi(x) = \pm \begin{cases}
    N \pi/2
     + \mbox{am} \left( \frac{x}{k\lambda }, \, k^2 \right), \mbox{ odd }N, \\
    (N-1) \pi /2
    + \mbox{am} \left( \frac{x}{k \lambda} +  K(k^2), \, k^2 \right), \mbox{ even }N,
    \end{cases}
\end{equation}
where $K(.)$ is the elliptic integral of the first kind and $\mbox{am}(.)$ is the Jacobi amplitude function, as EllipticK[.] and JacobiAmplitude[.] used in Wolfram Mathematica. The parameter $k$ ($0 \le k \le 1$) is solved self-consistently from the condition
$2 k K(k^2) = 1/w$.

\end{document}